\pdfoutput=1
\documentclass[sigconf, nonacm]{acmart}
\newcommand{\topic}[1]{\vspace{-3.5pt}\smallskip \smallskip \noindent{\bf #1.}}

\newcommand{\techreport}[1]{}

\newcommand{\rev}[1]{#1}

\theoremstyle{definition}

\newcommand{\shreya}[1]{\textcolor{orange}{[Shreya: #1]}}
\newcommand{\agp}[1]{\textcolor{purple}{[Aditya: #1]}}
\newcommand{\karl}[1]{\textcolor{red}{[Karl: #1]}}
\newcommand{\labib}[1]{\textcolor{teal}{[Labib: #1]}}

\newcommand{\gate}{\textsc{gate}\xspace}

\newcommand*{\medcup}{\mathbin{\scalebox{1.5}{\ensuremath{\cup}}}}%

\def\dashedrule#1#2#3{{%
\dimen1=#2 \divide\dimen1 by 2
\def\@ruledash{%
\rule{\dimen1}{0pt}%
\rule[0.5ex]{#1}{0.4pt}%
\rule{\dimen1}{0pt}}%
\count1=0
\loop%
\ifnum\count1<#3%
\advance\count1 by 1%
\@ruledash%
\repeat}}

\usepackage[autostyle, english = american]{csquotes}
\MakeOuterQuote{"}
\usepackage{subcaption}
\usepackage{natbib}
\usepackage{cleveref}
\usepackage{booktabs}
\usepackage{multirow}
\usepackage{array}
\usepackage{algpseudocode}
\usepackage{algorithm}
\usepackage{amsmath}
\usepackage{smartdiagram}
\usepackage{tikz}
\usepackage{xcolor}
\usepackage{xspace}
\usepackage{colortbl} 
\usetikzlibrary{positioning}
\usetikzlibrary{shapes.geometric}
\usetikzlibrary{shadows.blur}
\usetikzlibrary{shapes.symbols}
\usetikzlibrary{backgrounds}
\usepackage{pgfplots}
\usepgfplotslibrary{dateplot}
\usepackage{mathtools}
\usepackage{bbm}

\DeclareMathOperator*{\argmin}{arg\,min}

\makeatletter
\DeclareRobustCommand
  \myvdots{\vbox{\baselineskip4\p@ \lineskiplimit\z@
    \kern2\p@\hbox{.}\hbox{.}\hbox{.}}}
\makeatother
\AtBeginDocument{%
  \providecommand\BibTeX{{%
    \normalfont B\kern-0.5em{\scshape i\kern-0.25em b}\kern-0.8em\TeX}}}

\setcopyright{acmcopyright}
\copyrightyear{2018}
\acmYear{2018}
\acmDOI{XXXXXXX.XXXXXXX}

\acmConference[Conference acronym 'XX]{Make sure to enter the correct
  conference title from your rights confirmation emai}{June 03--05,
  2018}{Woodstock, NY}
\acmBooktitle{Woodstock '18: ACM Symposium on Neural Gaze Detection,
 June 03--05, 2018, Woodstock, NY} 
\acmPrice{15.00}
\acmISBN{978-1-4503-XXXX-X/18/06}

\makeatletter
\pgfmathdeclarefunction{invgauss}{2}{%
  \pgfmathln{#1}
  \pgfmathmultiply@{\pgfmathresult}{-2}%
  \pgfmathsqrt@{\pgfmathresult}%
  \let\@radius=\pgfmathresult%
  \pgfmathmultiply{6.28318531}{#2}
  \pgfmathdeg@{\pgfmathresult}%
  \pgfmathcos@{\pgfmathresult}%
  \pgfmathmultiply@{\pgfmathresult}{\@radius}%
}

\pgfmathdeclarefunction{randnormal}{0}{%
  \pgfmathrnd@
  \ifdim\pgfmathresult pt=0.0pt\relax%
    \def\pgfmathresult{0.00001}%
  \fi%
  \let\@tmp=\pgfmathresult%
  \pgfmathrnd@%
  \ifdim\pgfmathresult pt=0.0pt\relax%
    \def\pgfmathresult{0.00001}%
  \fi
  \pgfmathinvgauss@{\pgfmathresult}{\@tmp}%
}
\makeatother

\usepackage{array}
\newcolumntype{Y}{>{\setbox0=\hbox\bgroup}c<{\egroup}@{}}
\newcolumntype{L}[1]{>{\raggedright\let\newline\\\arraybackslash\hspace{0pt}}m{#1}}

\newif\ifhidecomments
\hidecommentsfalse

\newcommand\vldbdoi{XX.XX/XXX.XX}

\newcommand\vldbavailabilityurl{URL_TO_YOUR_ARTIFACTS}

\begin{document}

\title{Moving Fast With Broken Data}
\subtitle{\rev{Implementing an Automatic Data Validation System for ML Pipelines}}


\author{Shreya Shankar$^*$}
\affiliation{%
\authornote{Work performed during an internship at Meta}
\institution{UC Berkeley}{shreyashankar@berkeley.edu}
}

\author{Labib Fawaz}
\affiliation{%
\institution{Meta}{labibfawaz@meta.com}
}

\author{Karl Gyllstrom}
\affiliation{%
\institution{Meta}{gylls@meta.com}
}

\author{Aditya G. Parameswaran}
\affiliation{%
\institution{UC Berkeley}{adityagp@berkeley.edu}
}

\begin{abstract}
Machine learning (ML) models in production pipelines are frequently retrained on the latest partitions of large, continually-growing datasets. Due to engineering bugs, partitions in such datasets almost always have some corrupted features; thus, it's critical to detect data issues and block retraining before downstream ML model accuracy decreases. However, it's difficult to identify when a partition is corrupted \emph{enough} to block retraining. Blocking too often yields stale model snapshots in production; blocking too little yields broken model snapshots in production.

In this paper, we \rev{present an automatic data validation system for ML pipelines implemented at Meta.} We \rev{employ what we call} a Partition Summarization (PS) approach to data validation: each timestamp-based partition of data is summarized with data quality metrics, and summaries are compared to detect corrupted partitions. We describe how we can adapt PS for several data validation methods \rev{and compare their pros and cons. Since none of the methods by themselves met our requirements for high precision and recall in detecting corruptions, we devised \gate, our high-precision and recall data validation method. \gate gave a $2.1\times$ average improvement in precision over the baseline on a case study with Instagram's data.} Finally, we discuss lessons learned from implementing data validation for Meta's production ML pipelines.
\end{abstract}

\maketitle



\ifhidecomments
\renewcommand{\shreya}[1]{}
\renewcommand{\agp}[1]{}
\renewcommand{\karl}[1]{}
\renewcommand{\labib}[1]{}
\fi

\section{Introduction}
\label{sec:intro}
Errors in input data can negatively 
impact machine learning (ML) performance, 
motivating data validation for ML pipelines~\cite{breck, datavalexp, Biessmann2021AutomatedDV}. 
\citet{breck} state that 
``the importance of this problem is hard to overstate, 
especially for production pipelines,'' 
where ML models are frequently retrained 
on the latest partitions of data---including 
the underlying input features as well as the corresponding predictions~\cite{prodmlanalysisandopp}. \rev{For instance, suppose the audio doesn't work for the newest release of the Facebook app. The performance of a news feed ranking model might suffer, as audio-related features for users who updated their apps would be corrupted.
While an} obvious consequence of a corrupted partition \rev{(i.e., day)} 
of inputs is low model performance 
or accuracy for that partition, \rev{a} lesser-obvious and longer-lasting consequence 
of a corrupted partition 
is that models retrained 
on that partition will also be corrupted, 
spawning a cycle of errors 
that's difficult to recover from. \rev{In our example, once the audio is fixed, ML performance would actually diverge again because models would have been retrained on corrupted audio features (e.g., a binary indicator representing whether the sound is muted).}
In this paper, we therefore focus on the problem
of {\bf automatically validating
data to detect issues in production 
ML pipelines, {\em before} models are retrained.}

Detecting data errors in production ML pipelines 
before they cause downstream accuracy drops is hard \rev{at the scale of Meta}. 
Data partitions can be large (i.e., several petabytes), 
with tens of thousands of features 
(i.e., input data to the model).  
Given so many features, it is almost always the case that engineering bugs 
outside the model developers’ control corrupt some of them---especially when, at many organizations \rev{like Meta}, feature generation \rev{pipelines} are separated from \rev{calls to} ML models~\cite{opmlinterview, prodmlanalysisandopp}. Additionally, it's unclear what threshold of corruption results in a tangible impact on accuracy~\cite{datalifecyclechallenges}. ML models can actually tolerate some errors: for example, if an unimportant feature breaks and becomes null \rev{or garbage-}valued, the ML performance (i.e., prediction accuracy) usually remains the same. It's difficult to know \emph{how} broken a partition can be before significantly impacting performance. Moreover, this corruption tolerance is different for each pipeline, requiring pipelines to have on-call engineers that investigate whether production model snapshots are broken and should be rolled back to earlier model versions~\cite{opmlinterview}. A data validation system should alert \rev{Meta's on-call engineers} to gate, or block, the promotion of a model snapshot to production if necessary, rather than wait for the quality of subsequent predictions to degrade.

\topic{Prior \rev{data validation techniques don't solve our problem}} Work on ML data validation 
provides limited insights on how
to address the aforementioned challenges, either 
{\em (i)} \rev{being} overly reliant on manual input,
{\em (ii)} \rev{being} too coarse-grained, or
\rev{{\em (iii)} \rev{generating} too many false positive alerts.} 
As an example of {\em (i)},
\citet{automatinglargescale} 
propose a variety of 
general-purpose large-scale 
data validation techniques; however, these techniques 
aren't operationally scalable 
because they require \rev{our} engineers 
to enumerate and fine-tune constraints for each feature.
As an example of {\em (ii)},
\citet{breck} propose {\em schema validation}
techniques for production ML pipelines at Google.
These techniques check each tuple 
against a schema, 
consisting of type checks and loosely-defined bounds 
(e.g., non-negativity for an age-related feature). Schema validation is necessary but not sufficient; most ML pipelines have some schema validation but still experience many data corruption issues.
\citet{breck} also introduce a technique to
quantify differences in feature distributions, e.g., via Earth-Movers distance. 
However, 
 \rev{with regards to {\em (iii)}},
both \citet{breck} and \rev{our work find} 
that triggering alerts based on 
high difference measures yields too many false positive alerts, 
especially in datasets with many features. 
Moreover, most ML data validation 
methods incorrectly trigger alerts for \emph{expected} abnormalities. They validate the latest partition of data by
comparing it to only its previous partition or a single aggregation of historical partitions~\cite{deequ, breck, automatinglargescale, datasentinel, Ehrlinger2019ADT}. 
Doing so fails to account for expected anomalies that are tied to temporal patterns---\rev{for example, popular shopping days like Prime Day or Black Friday might temporarily inflate user engagement features and ad click-through rate. More commonly, user engagement fluctuates with typical seasonality, like day-vs-night or weekday-vs-weekend patterns.} Data validation methods should trigger alerts for \emph{unexpected} dataset corruptions, or those that actually warrant an engineer's attention.

\topic{Key insight \rev{for scale}: Partition Summarization} \rev{A simple idea that has been used in other data management contexts, such as in query optimization and approximate query processing, e.g.,~\cite{Moerkotte1998SmallMA, Li2018ApproximateQP}, is to store summaries of partitions for use in later computation. To identify dataset corruptions, we can compare a new partition's summary to a number of historical summaries to determine whether it is a true anomaly. For example, the partition corresponding to Monday is compared not just with Sunday, or a coarse-grained aggregate for the last week, but individual partitions for each day of the week---which includes both weekends and weekdays. We call this approach to ML data validation {\em Partition Summarization (PS)}.} To the best of our knowledge, \citet{Redyuk2021AutomatingDQ}'s  technique 
is the only prior work \rev{in data validation for ML} that
employs partition summarization. \rev{We implemented their technique as a baseline and found that it still triggered many false positive alerts, since} it is designed to alert general dataset errors, \rev{as opposed to} ML pipeline errors, \rev{as we will explain further in \Cref{sec:expts-discussion-baselinedrawback}.} \rev{{ Our main contribution is thus a general framework for adapting \emph{any} data validation for ML method to the PS setting.} We developed several adaptations, and since they did not meet our alert precision requirements, we additionally created \gate to achieve high precision for our workloads. Given the highly-correlated and high-dimensional nature of ML pipeline datasets, \gate employs the following steps: {\em (i)} clustering correlated features based on partition summaries, {\em (ii)} computing statistical measures of data quality for features, and {\em (iii)} monitoring cluster-wide aggregate statistics over time.}

\topic{Outline} \rev{In this paper, we give an overview of our recent effort to implement an automatic data validation system for ML pipelines at Meta.} Our system focuses on validating partitions of data for ML pipelines to detect issues before models are retrained. \rev{Our contributions are the following:}

\rev{\begin{itemize}
    \item We \rev{give an overview of our ML pipelines and} enumerate business requirements for an automatic data validation system (\Cref{sec:background}),
    \item We describe a general framework for adapting existing data validation methods to the Partition Summarization (PS) \rev{framework}, but these methods do not meet our requirements---namely, they still result in too many false positive alerts (\Cref{sec:problem,sec:solution}),
    \item We introduce \gate, our new, high-precision and high-recall automatic data validation technique that employs the PS framework (\Cref{sec:solution-gate}), and 
    \item We discuss takeaways from a case study on Instagram ML pipelines (\Cref{sec:expts}). We compare our data validation methods to \citet{Redyuk2021AutomatingDQ} and discuss best methods for different ML pipeline settings. Finally, we mention an example of using our system during an on-call rotation to debug an ML pipeline performance drop. 

\end{itemize}}

\section{Background}
\label{sec:background}
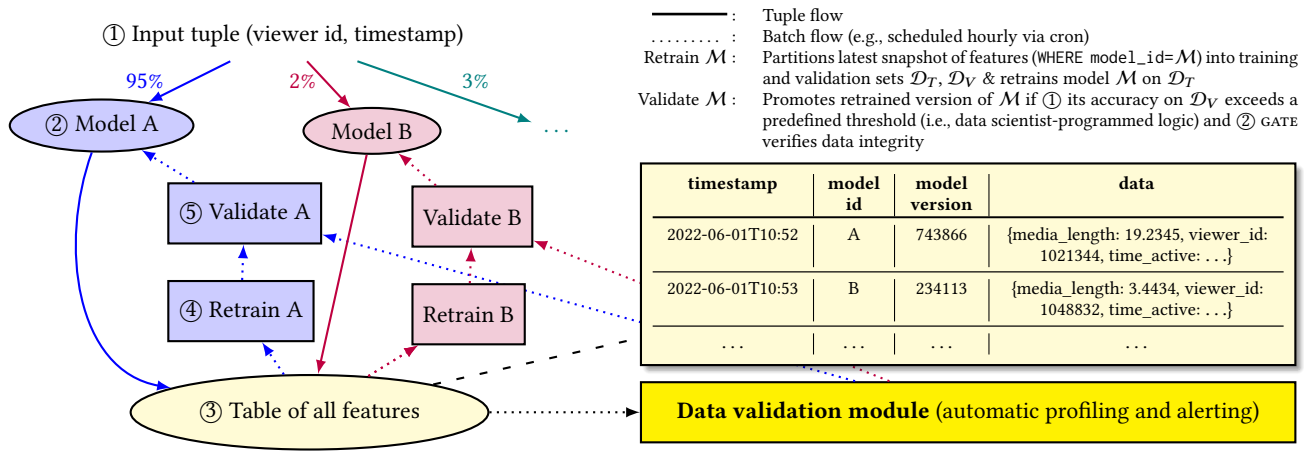
\begin{figure*}
    \centering
    {\begin{tikzpicture}[node distance={1.2cm}, thick, main/.style = {draw, rectangle, minimum size=0.8cm, inner sep=4pt}, model/.style = {draw, ellipse}, word/.style={inner sep=4pt}] 
        \node[word] (1) {\begin{tabular}{c} \textcircled{1} Input tuple (viewer id, timestamp) \end{tabular}}; 
        \node[model, fill=blue!20] (2) [below left=0.6cm and -1.2cm of 1] {\textcircled{2} Model A}; 
        \node[word, color=blue] (a) [below left=0cm and -1.3cm of 1] {95\%}; 
        \node[model, fill=purple!20] (3) [below right=0.7cm and -2.2cm of 1] {Model B}; 
        \node[word, color=purple] (b) [below right=0cm and -2.8cm of 1] {2\%}; 
       
        \node[model, inner sep=6pt, fill=yellow!20] (6) [below right=3.2cm and 0.2cm of 2] {\textcircled{3} Table of all features};
        \node[main, fill=blue!20] (7) [below right=0.5cm and 0cm of 2] {\textcircled{5} Validate A};
        \node[main, fill=blue!20] (8) [below right=1.8cm and 0cm of 2] {\textcircled{4} Retrain A};
        
        \node[main, minimum width=8.7cm, fill=yellow] (14) [right=2cm of 6] {{\bf \rev{Data validation module}} (automatic profiling and alerting)};
        \draw[-latex, dotted] (6) -- (14);
        \draw[-latex, dotted, color=blue] (14) -- (7);
        
        \node[main, fill=purple!20] (9) [below right=0.5cm and -0.13cm of 3] {Validate B};
        \draw[-latex, dotted, color=purple] (14) -- (9);
        \node[main, fill=purple!20] (10) [below right=1.8cm and -0.13cm of 3] {Retrain B};
        
        \node[word, color=teal] (11) [right=1.2cm of 3] {$\hdots$}; 
        \node[word, color=teal] (b) [below right=0cm and -0.5cm of 1] {3\%};

        \draw[-latex, color=blue] (1) -- (2);
        \draw[-latex, color=purple] (1) -- (3);
        \draw[-latex, color=teal] (1) -- (11);
        \draw[-latex, color=blue] (2) to[out=250,in=170] (6);
        \draw[-latex, dotted, color=blue] (6) -- (8);
        \draw[-latex, dotted, color=blue] (8) -- (7);
        \draw[-latex, dotted, color=blue] (7) -- (2);
        \draw[-latex, color=purple] (3) -- (6);
        \draw[-latex, dotted, color=purple] (6) -- (10);
        \draw[-latex, dotted, color=purple] (6) -- (10);
        \draw[-latex, dotted, color=purple] (10) -- (9);
        \draw[-latex, dotted, color=purple] (9) -- (3);

        \node[main, minimum width=8.7cm, fill=yellow!20, blur shadow={shadow blur steps=5,shadow blur extra rounding=1.3pt}] (12) [above=0.2cm of 14] {
        {\footnotesize
        \begin{tabular}
        {>{\centering\arraybackslash}p{0.1\textwidth}|%
        >{\centering\arraybackslash}p{0.04\textwidth}|%
        >{\centering\arraybackslash}p{0.05\textwidth}|%
        >{\centering\arraybackslash}p{0.2\textwidth}}
        \textbf{timestamp} &\textbf{model id} & \textbf{model version} & \textbf{data} \\
        \toprule
        2022-06-01T10:52 & A & 743866 & \{media\_length: 19.2345, viewer\_id: 1021344, time\_active: $\hdots$\} \\
        \midrule
        2022-06-01T10:53 & B & 234113 & \{media\_length: 3.4434, viewer\_id: 1048832, time\_active: $\hdots$\} \\
        \midrule
        $\hdots$ & $\hdots$ & $\hdots$ & $\hdots$ \\
        \end{tabular}}
        }; 
        
        \node[word, minimum width=8.7cm, inner sep=0pt] (13) [above=0.1cm of 12] {
        {\footnotesize
        \begin{tabular}
        {r>{\arraybackslash}p{0.4\textwidth}}
        \rule[2pt]{1cm}{1pt} : & Tuple flow \\
        \hbox to 1cm{\leaders\hbox to 3pt{\hss . \hss}\hfil} : & Batch flow (e.g., scheduled hourly via cron) \\
        Retrain $\mathcal{M}$ : & Partitions latest snapshot of features (\texttt{WHERE model\_id=$\mathcal{M}$}) into training and validation sets $\mathcal{D}_T, \mathcal{D}_V$ \& retrains model $\mathcal{M}$ on $\mathcal{D}_T$ \\
        Validate $\mathcal{M}$ : & Promotes retrained version of $\mathcal{M}$ if \textcircled{1} its accuracy on $\mathcal{D}_V$ exceeds a predefined threshold (i.e., data scientist-programmed logic) and \textcircled{2} \gate verifies data integrity \\
        \end{tabular}}
        };

    \draw[loosely dashed] (6) -- (12);
    \end{tikzpicture} }
    \caption{An ML pipeline for a single prediction task, consisting of a main model and many experimental models. If the latest snapshot of features is corrupted, even with model-specific validation, the resulting model version will have poor performance.} 
    \label{fig:pipeline}
\end{figure*}

In this section, we provide background on production ML pipelines, requirements for a practical data validation solution, and a categorization of existing data validation methods.

\subsection{ML Pipelines and Model Validation}
\label{sec:background-prodml}

A \emph{production ML pipeline} consists of one or more ML models that continuously produce predictions, in real-time, and are periodically retrained on the latest partition(s) of data. The boundary between training and deployment is \rev{blurry}: pipelines continually ingest input tuples (i.e., a row of feature values), run ML models in {\em inference mode} (i.e., don't change model parameters) for these tuples, and then later run ML models in {\em training mode} (i.e., change model parameters) for the latest batch of tuples.

\Cref{fig:pipeline} depicts an example production ML pipeline. Traffic to the pipeline is routed to either the main model model A or one of many experimental models created by data scientists (e.g., model B). Suppose an input tuple (\textcircled{1} in \Cref{fig:pipeline}) is routed to model A. Then, \textcircled{2} Model A will compute a prediction (e.g., probability of \rev{user} action). This prediction is returned to an end user or application and \textcircled{3} logged to a table with all features and predictions. Separately, every several hours or days (the cadence differs for different models and pipelines), \textcircled{4} retraining jobs are launched to retrain each model on the latest partition(s) of tuples---with the exception of a small hold-out set of tuples used for validation. Each retraining job verifies that the newly trained model snapshot achieves good performance (i.e., low loss) on the hold-out set, and if \textcircled{5} this \emph{model validation} step passes, the new model snapshot replaces its corresponding old production snapshot. ML engineers can include additional model-specific checks, like unit tests on specific tuples~\cite{modelassertions}. 

While model validation is necessary to prevent bad models from going to production, it is not sufficient. \citet{breck} discuss how bugs in code are a common source of production data errors and that the quality of ML predictions drastically suffers if features are corrupted. For example, a mobile app feedback logging mechanism might break, causing derived features to be null or $-1$ until an engineer fixes this logging mechanism. In many organizations \rev{like Meta}, the turnaround time to fix such bugs is often greater than the retraining cadence (i.e., engineers take more than a couple of hours to identify bugs, fix them, and push the change to production)---consequently, models might be retrained on corrupt or abnormally out-of-distribution data~\cite{opmlinterview}. Since the hold-out validation set is likely similarly corrupted, the retrained model will still pass the validation phase; however, the performance may drop significantly in production compared to historical performance. Thus, it is imperative to also perform \emph{data validation}, in addition to model validation, especially in the continual setting where data is constantly fed back to the model for retraining. 

\subsection{Data Validation Requirements \rev{at Meta}} \label{sec:background-dvreqs}

Our data validation goal is to determine whether there are enough ``invalid'' tuples in the latest partitions of training datasets to cause a downstream model performance drop if the retrained model snapshot is pushed to production. Trivially, we could \emph{gate}, or block, the promotion of all retrained models to production, but then the model snapshot in production would quickly get stale. Thus, it is important to \emph{precisely} identify when and how the data is corrupted. 

\rev{While the ML infrastructure team at Meta had already implemented schema validation and other constraint checking techniques described in~\citet{breck} and \citet{automatinglargescale}, these techniques did not precisely catch our ML failures. Moreover, ML engineers had to enumerate features and constraints for their pipelines, which was too hard to sustain over time as data changed and teams experienced natural turnover. As we started building a new data validation system, we first informally interviewed ML engineers at Meta} to collect requirements, \rev{summarized} in the following paragraphs.

\topic{High-precision and high-recall alerts without manual tuning} Engineers explicitly require a minimum recall (otherwise the system will be useless) and implicitly require a minimum precision (otherwise they will not pay attention to system alerts). \rev{Our original approach therefore encouraged engineers to} set alerts on completeness of features, i.e., the fraction of non-null values. If a feature's completeness dropped by more than 30\%, \rev{the on-call ML engineer} received a notification. However, this method produced too many false positive alerts, and \rev{ML engineers} thus silenced them. Some \rev{ML engineers} discovered an increase in precision when manually tuning the threshold, but this approach required significant effort. Eventually, \rev{ML engineers} abandoned this data validation method.


\topic{Scaling to many correlated features} Pipelines consist of datasets that can have tens of thousands of features, many of which are highly correlated and contribute unequally to overall ML model performance. A data validation solution that treats features independently and identically will trigger false positive alerts.  While some organizations may be able to weight data quality alerts on features by their importances~\cite{ribeiro2016model}, it's not generalizable and computationally feasible at \rev{the} scale \rev{of Meta}. Not all models have clearly defined feature importance scores (e.g., deep learning), and because models are frequently retrained, feature importances change often.

\topic{Scaling to many ML pipelines} Pipelines \rev{at Meta} are diverse, spanning different applications (e.g., recommendation, churn prediction, automated replies) and end-users (e.g., customers in different geographic regions). ML pipelines vary in their data ingestion and training dynamics---for example, some pipelines have higher probabilities of corrupted data than others, and some ML tasks are ``harder'' to solve than others (e.g., fraud detection is challenging due to class imbalance). Our data validation platform must support a variety of patterns in pipeline dynamics.

\topic{Ease of onboarding and use} When ML teams request data validation, we must roll out a solution immediately rather than collect annotations of ground-truth corruptions over weeks or months to tailor a validation system specific to their data or pipelines. Moreover, \rev{ML engineers are often} responsible for ML pipelines that consist of features that they don't create (due to organizational turnover). It's impractical to enumerate and carefully tune constraints and thresholds for features they may not have context about. Finally, alerts should be interpretable \rev{for engineers to act on}. Since an engineering bug typically breaks a single feature (e.g., a typo corrupts a string-valued feature), alerts should map to a broken feature or set of features~\cite{shankar2021towards}. As such, monitoring uninterpretable low-rank representations of features (e.g., from PCA) is not useful.

\subsection{Data Validation for ML\rev{: Existing Measures}}
\label{sec:background-existingdv}


Classical data validation and repair techniques focus on metrics that broadly correspond to correctness (e.g., not violating schemas~\cite{breck, Stonebraker2013DataCA, automaticschemamatching}), consistency (e.g., adhering to functional dependencies), completeness (i.e., maintaining a small fraction of null values~\cite{dataprofiling, Redyuk2021AutomatingDQ}), and statistical properties (e.g., not having outliers relative to means or medians~\cite{Biessmann2021AutomatedDV, krishnan2017boostclean}). A data validation \emph{method} takes some data quality statistic(s) and triggers an alert if some \rev{condition} is satisfied. We categorize existing data validation methods across three axes:



\begin{enumerate}
    \item Intra-feature: point vs.\ density measures of features
    \item Inter-feature: individual vs.\ groups, or joint distributions
    \item Temporal: single snapshot vs.\ over time
\end{enumerate}

We subsequently categorize more common data validation methods across the three axes and demonstrate a simple extension for any method to support the time axis.

\subsubsection{Intra-Feature Validation} Existing metrics to assess the distribution of a single feature can be decomposed to ``point'' measures, such as means, and ``density'' measures, such as histograms or empirical cumulative density function (eCDFs). The following statistics (spanning both point and density measures) are commonly used for data cleaning and validation: completeness, mean, standard deviation, number of unique values, number of frequent values, and the count of the most frequently-occurring value~\cite{Hellerstein2008QuantitativeDC, automatinglargescale}. Metrics that don't capture density might lead to false positive alerts; for example, a few outlier values for a feature can change the mean but not the overall ML model performance.

\subsubsection{Inter-Feature Validation} Most techniques monitor univariate distributions independently and identically, which can lead to false positive alerts in the ML setting because features are correlated and unequally important~\cite{breck, datavalexp,automatinglargescale,rabanser2019failing}. One way to extend techniques to multivariate settings \rev{is to concatenate data quality measures for all historical features into a vector and use model-based techniques (e.g., $k$-nearest neighbors) to compare the current vector against the historical vector~\cite{blazquez2021review}.  However, without some clustering, each feature will still be considered independently and identically.} A solution is to group features and \rev{compute} aggregate (e.g., average) statistics for each group~\cite{breck, datavalexp}, but it's impractical to ask each model developer or maintainer to manually group thousands of features. Dimensionality reduction techniques such as principal component analysis (PCA) can reduce the rank of the data~\cite{partridge1998fast}, or the number of columns to compute statistics for, but these columns are not interpretable---for example, if a PCA component is found to have an anomalous mean, it is not clear how to debug it. \rev{Moreover, clustering techniques applied to petabytes of data are computationally infeasible, motivating some sampling---but it is necessary to maintain the temporal nature of data in samples, as discussed next.}


\subsubsection{Temporal Validation}  Static lower and upper bounds on features will become stale over time, but we could track statistical measures over rolling weekly windows to construct \emph{dynamic bounds}. Suppose we compute statistical measure $Q$ daily---then at time $t$, a valid range for $Q\left(t\right)$ could be:

\begin{align*}
\frac{1}{7} \left(\sum_{i=1}^{7} Q\left(t-i\right) \right)  \pm z \times \sigma \left( \left[ Q\left(t-1\right), Q\left(t-2\right), \hdots, Q\left(t-7\right) \right] \right)
\end{align*}

where $z$ is a threshold number of standard deviations, and $\sigma$ is the standard deviation. ML-based anomaly detection methods typically \rev{acknowledge the existence of a} time axis without needing an extension~\cite{blazquez2021review}: for example, \citet{Redyuk2021AutomatingDQ} use a nearest-neighbors model to compare data quality statistics across time. \rev{However, this method does not weight the data points temporally; for example, the algorithm does not distinguish between yesterday's data quality statistics and data quality statistics from some arbitrary day a month ago.} Autoregressive models \rev{temporally weight data points by tracking} estimated moving averages of statistics~\cite{newbold1983arima}\rev{, but they don't work as well for multivariate distributions, especially not at our scale of thousands or tens of thousands of features}.

\topic{Looking Ahead} \rev{To correctly trigger alerts for corrupted data partitions, it's important for a data validation method to capture all three axes---intra-feature, inter-feature, and temporal.} Many existing data validation methods vary in how effectively they accomplish the goals of each axis. For example, a common method that performs poorly on all three axes is computing the mean of a feature---say, number of views for a photo---and rejecting a tuple if the number of views lies outside some predefined, static bounds placed around the mean number of views~\cite{breck, datavalexp,automatinglargescale}. \rev{Our partition summarization framework, which detail in \Cref{sec:solution}, employs both point and density-based statistical measures (intra-feature validation), normalized with respect to their moving averages (temporal validation). In \Cref{sec:solution-gate}, we introduce \gate, a technique that additionally performs inter-feature validation by cheaply clustering features and preserving a map from cluster back to original feature (unlike PCA).}

\section{Problem}
\label{sec:problem}

In this section, we introduce our problem statement and discuss evaluation metrics for data validation methods.

\subsection{Formalization}
\label{sec:problem-formalization}

Consider a dataset $\mathcal{D}$ of timestamp-ordered partitions\footnote{In this paper, each partition represents a day (i.e., 24 hours) of data, but other breakdowns are possible (e.g., a partition per hour).} $\mathcal{D} = \{D^{1},$ $D^{2}, \hdots, D^{t} \}$ where $\mathcal{D}$ has features (i.e., columns) $F_1, F_2, \hdots, F_n$, and $F_j^i$ represents the multiset of values for feature $j$ in the $i$th partition of data, i.e., $D^{i}$. At every timestamp $t$, an ML model is retrained on the partitions until that point, i.e.,  $\medcup_{i < t} D^{i}$. Although the absolute performance of the model can be measured by different metrics, such as accuracy or loss, we generically define an ML \emph{performance drop} as a degradation in the metric of choice compared to a rolling baseline, e.g., 5\% drop in accuracy compared to the 7-day rolling average accuracy. Typically, an ML engineer sets this definition and threshold. A data partition $D^i$ is \emph{corrupted} if a model retrained on $D^i$ experiences a performance drop at the following timestamp, i.e., $i + 1$. We denote $\mathcal{D}$'s corruptions with $y^i \in \{0, 1\}$, where $y^i = 1$ if and only if $D^i$ is corrupted.


We then define a {\em data validation method} to be a function that returns a score indicating the likelihood of a corruption for $D^t$, given historical partitions $D^1, \hdots, D^{t-1}$. More formally, a data validation method $f$ is defined as $ f\left(D^{t} \mid D^{1}, D^{2}, \hdots, D^{t-1} \right)$. If the arguments are clear from the context, we will use the short notation $f\left(t\right)$ for a data validation method.  Practically, $f\left(t\right)$ should be inexpensive to compute---especially at scale, but how do we evaluate its performance, i.e., \rev{if} $f\left(t\right) \approx y^t$? We discuss evaluation metrics next.

\subsection{Evaluation Metrics}
\label{sec:problem-evalmetrics}



Typically, data validation methods produce alerts, where an alert is triggered when $f\left(t\right)$ exceeds some threshold (e.g., 50\%). Given alert threshold $\tau$, we let $a_{\tau}^t = \mathbbm{I} \left[f \left(t \right) \geq \tau \right]$; so we want $a_\tau^t \approx y^t$---i.e., the alert should match the drop in data quality. We define precision and recall at $t$:
\begin{align*}
P(t, \tau) = \frac{\sum_{i=1}^t y^{i} \cdot a_{\tau}^i}{\sum_{i=1}^t  a_{\tau}^i} \quad
R(t, \tau) = \frac{\sum_{i=1}^t y^{i} \cdot a_{\tau}^i}{\sum_{i=1}^t y^{i}}
\end{align*}

A data validation method may generate at most $t$ distinct scores, \rev{one corresponding to each $f\left(i\right)$ such that $i \in \{1, \hdots, t\}$,} giving us at most $t$ thresholds $\tau_1 \hdots \tau_t$ to choose from. \rev{Specifically, at Meta we are interested in {\bf precision@0.9}, or the precision at the threshold $\tau$ that gives 90\% recall. This metric is of primary interest to practitioners, since a method that cannot recall most failures does not meet the bar for deployment. In addition to precision@0.9, we also consider another metric---{\bf average precision (AP)}---that takes into account the overall shape of the precision-recall (P-R) curve across thresholds~\cite{recall2004precision}.} AP is the weighted mean of precisions at each threshold, where the weight is the increase in recall from the previous threshold. \Cref{eq:ap} shows the definition of AP at $t$, if $P(t, \tau_i)$ and $R(t, \tau_i)$ are the precision and recall at the $i$th threshold $\tau_i$:
\begin{equation}\label{eq:ap}
AP\left(t\right) = \sum_{i=1}^t \left(R\left(t, \tau_i\right) - R\left(t, \tau_{i-1}\right)\right) P\left(t, \tau_i\right)
\end{equation}
AP \rev{captures the overall} predictive power of a method, or how the method performs when various thresholds of scores are chosen to fire alerts. As such, AP gives us a holistic, unbiased estimate of a method's performance. \rev{We leave a discussion of alternate evaluation metrics to \Cref{sec:related}.}

\section{Partition Summarization}
\label{sec:solution}

Existing data validation setups~\cite{breck, datavalexp, automatinglargescale} typically apply some statistical measure $Q$ (e.g., completeness, mean) to each feature $F_j$ at time $t$ and analyze how $Q \left(F_j^t\right)$ compares to $Q\left(F_j^{t-1}\right)$, the statistic for the previous day. Or, they compare $Q\left(F_j^t\right)$ to $Q\left( \cup_{i=1}^{t-1} F_j^i \right)$, i.e., the statistic computed on the union of all previous values\footnote{Here and elsewhere, when we use $\cup$, we are referring to the multiset union rather than the set union.}. Neither setup accounts for expected temporal patterns, e.g., weekends behaving differently compared to weekdays. Additionally, the latter setup loses partition-level granularity, which is important to preserve for identifying holidays.

The Partition Summarization (PS) approach to data validation \rev{that we introduce in this paper, extending \citet{Redyuk2021AutomatingDQ}'s method to any measure,}
involves first computing one
or more statistical summaries, 
$Q$, (e.g., mean) for each feature $F_j$ 
for each partition $D^i$ for $i \leq t$, 
i.e., $Q\left(F_j^i\right)$. 
Once these $Q$'s are computed and stored per partition,
we then combine them in various ways to compute $f\left(t\right)$.
Data validation methods can differ in the choice of statistics for partition summaries (i.e., 
$Q$) and how to combine these statistics to produce the overall
data quality scores (i.e., $f$). We describe a 
general adaptation of existing data validation methods to the PS setting in 
\Cref{sec:solution-adapting}, and we give specific adaptations in 
\Cref{sec:solutions-adaptations}. \rev{Finally, we discuss lessons learned from implementing these approaches in preliminary versions of our data validation system.}
 We place common notation in \Cref{tab:definitions}. 

\subsection{General Adaptation} \label{sec:solution-adapting}

As we described in Section~\ref{sec:problem}, in the PS framework,
we decompose data validation into two steps: one, where
we compute summaries $Q$ for features $F_1, F_2, \hdots, F_n$
across partitions $D^1, D^2, \hdots, D^t$;
and second, we combine these summaries to get
an overall measure of data quality $f(t)$ \rev{at each time step $t$}.
As a step towards computing $f(t)$, 
we aggregate each feature's difference 
between $Q$ and its rolling average $\overline{Q}$. 
In practice, we need to normalize 
the differences between $Q$ and $\overline{Q}$ before aggregating them.
Formally, given $Q$ and $F_i$, we define the rolling average $\overline{Q}$ over the last $k$ days as:
\begin{align}
\overline{Q} \left(F_i, t\right) &= \frac{1}{k} \sum_{j=1}^k Q\left(F_i^{t-j}\right)   
\label{eq:rollingQ}  &
\end{align}
Once we have rolling average $\overline{Q}$ per feature, 
we can normalize the $Q$s per feature as $\widetilde{Q}$.
The normalized $\widetilde{Q}$s can be computed using different normalization techniques, such as percent difference (PD) or $z$-score:
\begin{align}
\widetilde{Q}_{PD} \left(F_i, t\right) &= \frac{Q \left(F_i^t \right) - \overline{Q} \left(F_i, t \right)}{\overline{Q} \left(F_i, t \right)}  
\label{eq:percentdifferencenorm} &\\
\widetilde{Q}_{z} \left(F_i, t\right) &= \frac{Q \left(F_i^t \right) - \overline{Q} \left(F_i, t \right)}{\sigma \left(\left[Q\left(F_i^{t-1}\right), \hdots, Q\left(F_i^{t-k}\right) \right] \right)}  
\label{eq:zscorenorm} 
\end{align}
where $\sigma$ is a function that calculates the standard deviation. Finally, $f$ can be computed by aggregating the $\widetilde{Q}$'s for each feature, producing a scalar score:
\begin{align*}
f \left(t \right) &= \frac{1}{n} \sum_{i=1}^{n} \widetilde{Q} \left(F_i, t\right) 
\end{align*}

We can trigger an alert if $f \left(t\right)$ exceeds a threshold---which can be determined by keeping track of $f$ for a few timestamps and computing if the current value is an outlier. In practice, any simple scalar outlier detection mechanism can be used (e.g., multiple standard deviations greater than the mean, greater than 95th percentile). Since there is no ``training set'' for our approach, the initial time steps may incorrectly trigger alerts. However, an alert threshold chosen this way is robust to temporal variation (i.e., a percentile threshold will maintain its meaning and significance over time). \rev{In the following, we slightly abuse notation to adapt our approach to various data quality metrics: briefly, $\overline Q$ indicates a rolling average measure, $\widetilde Q$ indicates a normalized version (using $\overline Q$) for the current time step relative to others, and $f \left(t\right)$ aggregates $\widetilde Q$ across features.}

\begin{table}
    \centering
    \footnotesize
    \begin{tabular}{|r|L{6.8cm}|}
     \toprule
    {\bf Symbol} & {\bf Description} \\
    \toprule
        $D^{t}$ & Partition of data at timestamp $t$ \\
        \midrule
        $y$ & Binary labels representing whether partitions are corrupted, $y^t = 1$ if $D^t$ is corrupted and $0$ otherwise \\
        \midrule
        $F$ & A feature (i.e., column) in the dataset; $F_j$ represents the $j$th feature \\
        \midrule
       $f$  & Data validation method, $f: D^{t} \to \mathbb{R}$ \\
       \midrule
       $Q$ & Data quality statistic (e.g., mean, completeness), $Q: F^t \to \mathbb{R}$ \\
        \midrule
       $\overline{Q}$ & Average of the most recent statistical measures $Q$ for a feature, defined in \Cref{eq:rollingQ} \\ 
       \midrule
       $\widetilde{Q}_{PD}$ & Percent difference between $Q$ and its rolling average $\overline{Q}$, defined in \Cref{eq:percentdifferencenorm}\\
       \midrule 
       $\widetilde{Q}_{z}$ & $z$-score difference (i.e., number of standard deviations) between $Q$ and its rolling average $\overline{Q}$, defined in \Cref{eq:zscorenorm}\\
       \midrule
       $C$ & Completeness of a feature for a partition (i.e., the fraction of non-null values), $C: F^t \to \left[0, 1\right]$ \\
       \midrule
       $G^t$  & \gate's clustering assignment at time $t$, where $G^t\left(F\right)$ represents $F$'s cluster and $|G^t| = v$ is the number of distinct clusters \\
    \bottomrule
    \end{tabular}
    \caption{Notation Table}
    \label{tab:definitions}
\end{table}

\subsection{Adaptations of Existing Approaches}
\label{sec:solutions-adaptations}

We pick \rev{three categories of data quality measures to adapt to the PS setting: (1) monitoring the percent drop of completeness for each feature, (2) monitoring $z$-scores of any scalar statistic (e.g., completeness, mean) for each feature, and (3) monitoring $p$-values from two-sample statistical tests (e.g., Kolmogorov-Smirnov~\cite{massey1951kolmogorov}) for each feature.} For each of the methods, we set $k=7$, or one week, in computing $\overline{Q}$ and $\widetilde{Q}$. \rev{We chose $k=7$ because of typical organizational considerations (e.g., on-call rotation lengths are based on the week, meetings are often weekly), but this parameter can vary.} 

\subsubsection{$\delta$-Completeness Drop}\label{sec:solution-adaptation-completeness} The completeness drop method creates a score at time $t$ based on the number of features that have experienced a completeness drop (i.e., $\widetilde{C}$) $\geq \delta$ with respect to their rolling 7-day average completeness. The score is weighted by each feature's rolling average completeness (i.e., $\overline{C})$. \Cref{eq:canomaly} shows the $\delta$-completeness drop score for time $t$:
\begin{flalign}
\widetilde{C}_{PD}\left(F_i, t, \delta\right) &= \begin{cases} \widetilde{C}_{PD}\left(F_i, t\right) \times \overline{C}\left(F_i, t\right) & \text{if } \widetilde{C}_{PD}\left(F_i, t\right) \geq \delta \\ 0 & \text{otherwise} \end{cases} \nonumber\\
f_C\left(t, \delta\right) &= \frac{1}{n} \sum_{i=1}^n  \left| \widetilde{C}_{PD}\left(F_i, t, \delta\right) \right|  \label{eq:canomaly}
\end{flalign}
Note that we weight each feature that exceeds $\delta$ by its rolling average $\overline{C}$. This is because we want a feature to contribute significantly to a $f_C$ score when it  has a large change from its historical rolling average \emph{and} typically has a large completeness. A high $f_C$ score at time $t$ means that several high-completeness features now have low completeness measures.


\subsubsection{$z$-Score Anomaly Detection}\label{sec:solution-adaptation-zscore} In this method, we fix a scalar statistic (e.g., completeness or mean) and create a score at time $t$ based on the fraction of features that have a $z$-score outside some cutoff $\tau$. $z$-scores are computed using 7-day rolling means and standard deviations of the statistic. Given statistic $Q$, the anomaly detection score for time $t$ is shown in \Cref{eq:zscoreanomaly}:
\begin{flalign}
\widetilde{Q}_z\left(F_i, t, \tau\right) &= \begin{cases} \left| \widetilde{Q}_z\left(F_i, t\right) \right| & \text{if } \left| \widetilde{Q}_z\left(F_i, t\right) \right| \geq \tau \\ 0 & \text{otherwise}\end{cases}  \nonumber\\
f_z\left(t, \tau\right) &= \frac{1}{n} \sum_{i=1}^n \widetilde{Q}_z\left(F_i, t, \tau\right) \label{eq:zscoreanomaly}
\end{flalign}
Intuitively, we zero the contributing $z$-score for features with small $z$-scores (i.e., $< \tau$) because, given the tens of thousands of features we must monitor, many features are likely to have normal $z$-scores (i.e., less than 3). We want the ``most anomalous'' features to significantly alter the resulting $f_z$ score---for example, one feature's $z$-score being 15 is more alarm-worthy than 15 features each having a $z$-score of 1. Before adding the intermediate $\widetilde{Q}_z(F_i, t, \tau)$ step (i.e., the indicator for $\geq \tau$), this method performed very poorly.

\subsubsection{Two-Sample Statistical Tests}\label{sec:solution-adaptation-twosamp} To consider entire \emph{distributions} of each feature, we evaluated $p$-values based on three different statistical measures: Kolmogorov-Smirnov (KS), Wasserstein-1 ($\mathrm{wass}$) or Earth-Mover's Distance, and DTS~\cite{massey1951kolmogorov,vallender1974calculation,Dowd2020ANE}. At a high level, the KS measure computes the largest difference between two eCDFs at a single value, the Wasserstein-1 measure computes the entire area of difference between two eCDFs, and the DTS measure weights the Wasserstein-1 measure by variance of the combined eCDFs (denoted as $\hat D$ in \Cref{eq:dts}). For our \rev{implementation}, the eCDFs consist of 99 percentiles, or the 1st percentile to 99th percentile. If ${\bf eCDF}\left(F_i, t\right) \in \mathbb{R}^{99}$ represents an eCDF of feature $F_i$ at time $t$, the two-sample test statistics are shown in \Cref{eq:ks,eq:wass,eq:dts}:

\begin{flalign}
\mathbf{diff}\left(F_i, t\right) &= \left|{\bf eCDF}\left(F_i, t\right) - \frac{1}{7}\left(\sum_{j=1}^7 {\bf eCDF}\left(F_i, t-j\right) \right)\right| \nonumber\\ 
h_{\mathrm{KS}}\left(F_i, t\right) &= \max \mathbf{diff}\left(F_i, t\right)  \label{eq:ks} \\
h_{\mathrm{wass}}\left(F_i, t\right) &= \sum \mathbf{diff}\left(F_i, t\right) \label{eq:wass} \\
\hat D\left(F_i, t\right) &= \sigma \left({\bf eCDF}\left(F_i, t\right),  \frac{1}{7}\left(\sum_{j=1}^7 {\bf eCDF}\left(F_i, t-j\right) \right) \right) ^2  \nonumber\\
h_{\mathrm{DTS}}\left(F_i, t\right) &= \frac{\sum \mathbf{diff}\left(F_i, t\right) }{\hat D\left(F_i, t\right) \times \left(1 - \hat D\left(F_i, t\right) \right)}  \label{eq:dts}
\end{flalign}

Note that $\mathbf{diff}\left(F_i, t\right) \in \mathbb{R}^{99}$ represents an element-wise absolute value difference between the eCDF at time $t$ and the 7-day rolling average eCDF. To convert a statistical measure to a $p$-value, we bootstrap estimates of the statistical measure while randomly partitioning the current and historical average eCDFs and compute the fraction of bootstrapped values that don't exceed the original measure.

To convert the $p$-values to a scalar score for $t$, we compute the fraction of features with a $p$-value smaller than a significance level $\alpha$, typically set to 0.05.  If $p\left(F_i, t\right)$ is a $p$-value from a two-sample statistical test measure at time $t$ for feature $F_i$, the resulting score for this measure at time $t$ is shown in \Cref{eq:2sampscore}:

\begin{flalign}
\widetilde Q_{\mathrm{twosamp}} \left(F_i, t, \alpha\right) &= \mathbbm{I} \left[ p\left(F_i, t\right) \leq \alpha\right]  \nonumber\\
f_{\mathrm{twosamp}} \left(t \right) &=  \frac{1}{n} \sum_{1=1}^n \widetilde Q_{\mathrm{twosamp}} \left(F_i, t, \alpha\right) \label{eq:2sampscore}
\end{flalign}

\subsection{\rev{Lessons Learned}}
\label{sec:solution-lessons}

\rev{While the aforementioned adaptations of existing data validation techniques do have their benefits (as we will see in \Cref{sec:expts-discussion-bestmethods}), they did not meet our precision@0.9 requirements in preliminary collaborations with ML engineers at Meta. Here, we discuss high-level takeaways for each method.}

\subsubsection{$\delta$-Completeness Drop} Recall that $\delta$ represents the threshold for how much the completeness for any feature should drop from a rolling average to trigger an alert. For simplicity, engineers at Meta preferred to set the same $\delta$ for every feature, leading to many false positive alerts. Moreover, engineers spent several months tuning $\delta$s, and we found that the ideal $\delta$ varies significantly between datasets, \rev{motivating a method that did not require so much hand-tuning.}



\subsubsection{$z$-Score Anomaly Detection} Recall, from \Cref{sec:solution-adaptation-zscore}, that this method can be performed for any scalar statistic $Q$ (e.g., mean, completeness). We found completeness to give the best precision@0.9 and AP, and we hypothesize that this is because completeness results in fewer false positives than other statistics (e.g., a large, unexpected fraction of nulls is more likely to mess up downstream ML model performance than a large mean, which could have been influenced by a small fraction of outliers). Like the $\delta$-completeness drop method, this $z$-score anomaly detection method still yielded many false positives when requiring high recall.



\subsubsection{Two-Sample Statistical Tests} While literature seems to indicate that the KS test is a popular practical ML drift detection technique~\cite{dos2016fast, lu2014concept, rabanser2019failing}, we found that the KS test also yielded a large number of false positives. \rev{Deeper introspection reveals that the KS test is actually inappropriate for our problem.} As described in \Cref{sec:solution-adaptation-twosamp}, the KS test only considers the maximum distance (i.e., $L^{\infty}$) between two eCDFs. This is bad in the ML setting, as demonstrated by the following scenario: imagine two eCDFs (current and historical) that are mainly the same, except 1\% of the tuples are outliers in the current eCDF. Since only a small fraction of tuples are corrupted, the downstream ML model performance may not visibly degrade---but the KS test statistic may be large! As a result, we found that statistical tests based on the $L^1$ distance between the eCDFs (e.g., Wasserstein-1, DTS) largely outperformed the KS test, as corroborated by experiments in \citet{breck}. Wasserstein-1 performed the best. Moreover, because of need to bootstrap to generate $p$-values, the two-sample tests can run multiple orders of magnitude slower than other methods. We found that such methods could not scale to our high-dimensional datasets, some of which can have hundreds of thousands of columns.

\section{\gate}
\label{sec:solution-gate}
\begin{figure*}
    \centering
    \resizebox{\linewidth}{!}{\begin{tikzpicture}
    
     \pgfmathsetseed{1} 
     
     \draw[red, densely dashed, line width=2pt] (-0.5,-1) rectangle (15,4);
     
     \node[black, align=center] (fullpartition) at (-2.8, 0.7) {\footnotesize \begin{tabular}{c|c|c|c}
            f1 & f2 & f3 & $\hdots$ \\
            \toprule
            $\myvdots$ & $\myvdots$ & $\myvdots$ & $\myvdots$
        \end{tabular}};
    
    \draw [-latex] (-2.8, 1.3) -- (-2.8, 1.9);
    
    \node[black, align=center] (summary) at (-2.8, 2.8) {\footnotesize \begin{tabular}{c|c|c|c}
            feature & $\mu$ & $\sigma$ & $\hdots$ \\
            \toprule
            f1 & \multirow{2}{*}{$\myvdots$}  & \multirow{2}{*}{$\myvdots$}& \multirow{2}{*}{$\myvdots$} \\
            f2 & & & \\
            $\myvdots$ & $\myvdots$ & $\myvdots$ & $\myvdots$
        \end{tabular}};
    
     \draw [line width=2pt, -latex](-1.1, 2.8) -- (-0.1, 2.8);
    
    \node[black, align=center] (label0) at (-2.8, -0.5) {{\begin{tabular}{c} {\bf Partition} \\ {\bf Summarization} \end{tabular}}};
     

        \foreach \i in {0,...,10} {
          \fill [opacity=0.8] (1.3+randnormal/3, 2.8+randnormal/4) circle (0.05);
          \fill [opacity=0.8] (3+randnormal/5, 2.9+randnormal/5) circle (0.05);
          \fill [opacity=0.8] (3.1+randnormal/4, 1+randnormal/5) circle (0.05);
          \fill [opacity=0.8] (1.2+randnormal/5, 0.8+randnormal/4) circle (0.05);
        }

        \filldraw [fill=red, fill opacity=0.1, draw=red, densely dashed]
            plot [smooth cycle, tension=1] coordinates {(0.5,3.2) (2,3) (1,2)};
        \node[red] (g1) at (0.3,2.3) {$G_1$};

        \draw [fill=blue, fill opacity=0.1, draw=blue, densely dashed]
            plot [smooth cycle, tension=1] coordinates {(2.5,2.8) (3.7,2.5) (3,3.5)};
        \node[blue] (g2) at (2.5,2.4) {$G_2$};
        
        \draw [fill=teal, fill opacity=0.1, draw=teal, densely dashed]
            plot [smooth cycle, tension=1] coordinates {(1,0.1) (0.8, 1) (1.5, 1.3) (1.8, 1)};
        \node[teal] (g3) at (0.5,1) {$G_3$};

        \draw [fill=orange, fill opacity=0.1, draw=orange, densely dashed]
            plot [smooth cycle, tension=1] coordinates {(2.1, 0.7) (2.6, 1.2) (3, 1.6) (3.8, 1) (3, 0.5)};
        \node[orange] (g4) at (3.5,0.3) {$G_4$};
    
        \node[black] (label1) at (2, -0.5) {{\begin{tabular}{c} {\bf Decorrelation} \\ (Inter-feature validation) \end{tabular}}};
    
    \draw [line width=2pt, -latex](4.25, 1.5) -- (5.25, 1.5);
    
    \filldraw[fill=orange!10, draw=black] (5.5,0.45) rectangle (8.6,3.15);
    \filldraw[fill=teal!10, draw=black] (5.65,0.3) rectangle (8.75,3);
    \filldraw[fill=blue!10, draw=black] (5.8,0.15) rectangle (8.9,2.85);
    \filldraw[fill=red!10, draw=black] (5.95,0.0) rectangle (9.05,2.7);
        
        \node[black] (anomalymatrix) at (7.5, 1.3) {\begin{tabular}{r>{\arraybackslash}r}
            completeness & 0.8 \\
            $\mu$:& 0.12 \\
            $\sigma$:& 0.3 \\
            \# unique vals:& 2.2 \\
            top frequency:& -1.3 \\
            Wass-1:& -1.3 \\
        \end{tabular}};
        
        \node[black] (label2) at (7.5, -0.5) {{\begin{tabular}{c} {\bf Anomaly Matrix Creation} \\ (Intra-feature validation) \end{tabular}}};
        
    \draw [line width=2pt, -latex](9.3, 1.5) -- (10.3, 1.5);
    
        \draw [-latex](10.5, 0.3) -- (14.5, 0.3);
    
    \node (vector1) at (11,2.1) {$\footnotesize
    \begin{array}{l}\\
            \left(
                \begin{array}{c}
                    \cellcolor{red!10}0.4 \\
                    \cellcolor{red!10}\myvdots \\
                    \cellcolor{blue!10}-0.4 \\
                    \cellcolor{blue!10}\myvdots \\
                    \cellcolor{teal!10}0.2 \\
                    \cellcolor{teal!10}\myvdots \\
                    \cellcolor{orange!10}0.6 \\
                    \cellcolor{orange!10}\myvdots
                \end{array}
            \right)
        \end{array}
    $};
    \draw (11, 0.2) -- (11, 0.4);
    \node[black] (tick1) at (11, 0.1) {$t-3$};
    
    \node (vector2) at (12.4,2.1) {$\footnotesize
    \begin{array}{l}\\
            \left(
                \begin{array}{c}
                    \cellcolor{red!10}0.6 \\
                    \cellcolor{red!10}\myvdots \\
                    \cellcolor{blue!10}-0.7 \\
                    \cellcolor{blue!10}\myvdots \\
                    \cellcolor{teal!10}0.3 \\
                    \cellcolor{teal!10}\myvdots \\
                    \cellcolor{orange!10}0.2 \\
                    \cellcolor{orange!10}\myvdots
                \end{array}
            \right)
        \end{array}
    $};
    \draw (12.4, 0.2) -- (12.4, 0.4);
    \node[black] (tick2) at (12.4, 0.1) {$t-2$};
    
    \node (vector3) at (13.8,2.1) {$\footnotesize
    \begin{array}{l}\\
            \left(
                \begin{array}{c}
                    \cellcolor{red!10}0.8 \\
                    \cellcolor{red!10}\myvdots \\
                    \cellcolor{blue!10}-0.9 \\
                    \cellcolor{blue!10}\myvdots \\
                    \cellcolor{teal!50}3.3 \\
                    \cellcolor{teal!10}\myvdots \\
                    \cellcolor{orange!10}0.4 \\
                    \cellcolor{orange!10}\myvdots
                \end{array}
            \right)
        \end{array}
    $};
    \draw (13.8, 0.2) -- (13.8, 0.4);
    \node[black] (tick3) at (13.8, 0.1) {$t-1$};
    
        \node[black] (label3) at (12.5, -0.5) {{\begin{tabular}{c} {\bf Alert Generation} \\ (Temporal validation) \end{tabular}}};
    
    \draw [line width=2pt, -latex](14.5, 1.8) -- (15.5, 1.8);
    \filldraw[fill=teal!10, draw=black] (15.7,0) rectangle (19.5,2.5);
    
        \node[black] (anomalymatrix2) at (17.5, 1.3) {\footnotesize \begin{tabular}{c|c|c}
            feature & completeness & $\hdots$ \\
            \toprule
            f1 & 0.1  & \multirow{2}{*}{$\myvdots$} \\
            f2 & \cellcolor{teal!50}{6.5} & \\
            $\myvdots$ & $\myvdots$ & $\myvdots$ \\
            $\myvdots$ & $\myvdots$ & $\myvdots$ 
        \end{tabular}};
        
        \node[black] (label4) at (17.6, -0.5) {{\begin{tabular}{c} {\bf Feature Drill-Down} \\ (Debugging) \end{tabular}}};
        
    \draw [line width=2pt, -latex](-1.1, 3.8) -- (7.2, 3.8) -- (7.2, 3.2);
        
    \end{tikzpicture}}
    \caption{Architecture of \rev{our} Partition Summarization (PS) approach to data validation for ML. The \textcolor{red}{red} box represents a \rev{general} data validation \rev{technique. We found that existing techniques did not achieve high precision and recall in identifying our ML performance drops---mainly, there were too many false positive alerts. Our technique, \gate, clusters correlated features and triggers an alert if an entire group's summary statistics are anomalous.} }
    \label{fig:gate}
\end{figure*}
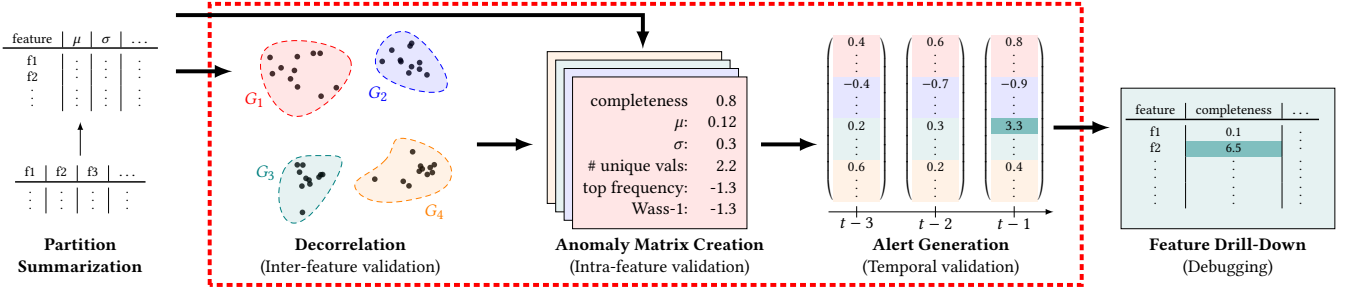


\rev{In this section, we describe \gate, our technique that results in fewer false positives than the methods from \Cref{sec:solution-adapting}, while being a bit more involved to implement.}

\subsection{Drawbacks of Existing PS Adaptations}

\rev{In designing \gate (\Cref{fig:gate}), we identified several drawbacks of off-the-shelf adaptations, described in \Cref{sec:solution-lessons}. First, they treat features as independently and uniformly important to the ML prediction. This is not true for our ML pipelines, as discussed in \Cref{sec:intro}. Clustering is thus crucial to reduce the number of false positive alerts in high-dimensional datasets, but it's not easy to cluster at scale and map back to original features for debugging. A solution we employ is to simply apply spectral clustering to covariance matrices of partition summaries, which runs quickly and preserves original features. Second, previous methods do not capture the entire density of feature distributions, which is usually missed with ``point'' statistics like the mean. As a solution, our summaries within \gate include a variety of statistics, including differences between cumulative density functions across timestamps. Third, it's hard to ensure that statistics are robust to expected temporal variation: for example, monotonically increasing features (e.g., total number of views) change over time, but we don't want to flag these as errors. As a solution, we normalize statistics to be $z$-scores, computed with respect to recent partition summaries, before comparing them.}

\subsection{Components of \gate}
\label{sec:solution-gate-overview}
 
To achieve high precision and recall, our technique, \gate, can be broken down into following components (as depicted in \Cref{fig:gate}):

\begin{enumerate}
    \item \textbf{Decorrelation:} To address inter-feature validation, we apply spectral decomposition to correlations between features. The decorrelation step accepts input partitions $D^{i}$ where $i \leq t$ and outputs $G^t \in  \mathbb{R}^n$, where each feature is represented by only one cluster. We denote the number of unique clusters $|G^t| = v$. For speed and scalability, we apply this clustering method to synopses or summary statistics instead of entire underlying datasets. (\Cref{sec:solution-decorrelation})
    \item \textbf{Anomaly matrix creation:} To address intra-feature validation, we compute 6 statistical measures for each feature and derive their $z$-scores for normalization purposes. We use $G^t \in \mathbb{R}^n$ to cluster and average the $z$-scores. This step returns $X\left(t\right) \in \mathbb{R}^{t \times 6v}$, where each row in $X\left(t\right)$ represents a measure of data quality for that timestamp. (\Cref{sec:solution-amc})
    \item \textbf{Alert generation:} To address temporal validation, we compute the average distance from ${X}_t\left(t\right)$ (the $t$th row in $X$) to a fraction of the closest previous rows and trigger an alert if the average distance exceeds some threshold. We can filter on recent timestamps if desired. This alert generation step accepts $X\left(t\right) \in \mathbb{R}^{t \times 6v}$ from the previous step, or the concatenated anomaly matrices from previous partitions, and outputs a corruption score. (\Cref{sec:solution-alertgen})
    \item \textbf{(Optional) Feature drill-down:} While this step is not critical for the alerting mechanism, ML engineers want to understand why an alert was triggered. We find the most anomalous clusters and ``drill-down'' into their features and corresponding anomalous statistics. (\Cref{sec:solution-drilldown})
\end{enumerate}

We subsequently discuss the steps in more detail.

\subsubsection{Decorrelation}
\label{sec:solution-decorrelation}

Since datasets can have tens of thousands of features, many of which are correlated and some of which do not contribute significantly to downstream ML model performance, monitoring features independently and with equal weight is likely to trigger false positive alerts in practice. To decrease the likelihood of a false-positive alert, we found it crucial to cluster features and monitor aggregates over resulting groups instead of individual features. To avoid leakage, we performed the clustering on approximately one month of historical data---such that our method would not ``peek into'' future data to identify correlations between features. Each partition of data can be several petabytes, so clustering raw data is impractical. So, in this clustering step, we clustered information from the partition summaries, not the raw data itself~\cite{birch}.

To perform the clustering, we leveraged a graph theoretic approach: nodes were features, and edge distances corresponded to the correlations between endpoint nodes (i.e., features)~\cite{white2005spectral}. First, for each of our 6 summary statistics (described further in~\Cref{sec:solution-amc}), we computed a covariance matrix across features using our summary table. To determine the number of clusters $|G^t|$, we ran PCA on each covariance matrix to determine the number of components $v_1, \hdots, v_6$ that would explain 95\% of variance. We set $|G^t| = \max v_1, \hdots, v_6$. Then we summed the absolute values of the covariance matrices to yield the graph edge matrix. Finally, we ran a spectral clustering algorithm to partition the graph into cluster assignments $G^t$, where each feature corresponded to one cluster. We denote the cluster for a feature $F$ as $G^t\left(F\right)$. 

There are a number of clustering methods in the literature~\cite{xu2005survey}; here, we picked spectral clustering for its simplicity and application to the graph setting. Since we wanted to cluster features (i.e., columns), instead of tuples, we simply constructed a graph of features as nodes, with edge weights being their correlations. Applying the spectral clustering algorithm could then identify groups of similar features, based on their correlations. Spectral clustering is ideal for nonconvex datasets but has some drawbacks: it doesn't work with noisy data and has a high runtime complexity~\cite{scar}. However, our technique doesn't face these problems because our summaries are aggregations and are thus less likely to be noisy, and the size of the covariance matrix (i.e., graph) is based on the number of features, not the number of tuples in the dataset. The size of the graph also doesn't depend on the number of partitions, so the clustering runtime will stay the same over time. \techreport{A full description of the algorithm is shown in \Cref{sec:app-algos}.}

\subsubsection{Anomaly Matrix Creation}
\label{sec:solution-amc}

The goal of this step is to turn $D^{t}$ into a vector of statistics to compare against previous partitions' vectors---essentially creating an matrix of statistics representing all $D^{1}, D^{2}, \hdots, D^{t}$. For our algorithm, we used the following six statistics: (1) completeness, (2) mean, (3) standard deviation, (4) number of unique values, approximated via sketches (5) top frequency (i.e.,  count of the most frequently-occurring value divided by total count) and (6) Wasserstein-1 (i.e., Earth-Mover's) distance between consecutive partitions' eCDFs. To normalize statistics amongst features and time, we turn each statistic into a $z$-score. Then, we reduce dimensionality by averaging $z$-scores across clusters. The clustering step (\Cref{sec:solution-decorrelation}) weights features unequally---based on correlations---for the alert generation step (\Cref{sec:solution-alertgen}).



Given clustering assignment $G^t$ from~\Cref{sec:solution-decorrelation}, $|G^t| = v$ distinct clusters, and statistics $Q_1, Q_2, \hdots, Q_6$, at time $t$, we average the normalized statistics $\widetilde{Q}_z$ within clusters to get \Cref{eq:gatex}:

\begin{flalign}
\widetilde{\bf Q}_z\left(F_i, t\right) &= \left[\left|\widetilde{Q_1}_z \left(F_i, t \right)\right|, \hdots, \left|\widetilde{Q_6}_z \left(F_i, t \right)\right|\right] &\in\;& \mathbb{R}_{+}^{6} &\nonumber\\
\widetilde{\bf Q}_G\left(j, t \right) &= \frac{\sum_{i=1}^n   \mathbbm{I}  \left[G^t\left(F_i\right) = j \right] \cdot \widetilde{\bf Q}_z \left(F_i, t\right)}{\sum_{i=1}^n \mathbbm{I} \left[G^t\left(F_i\right) = j \right] }  &\in\;& \mathbb{R}_{+}^{6} &\nonumber \\
\mathbf{x} \left(t\right) &= \left[\widetilde{\bf Q}_G\left(1, t \right) \; \widetilde{\bf Q}_G\left(2, t \right) \; \hdots \; \widetilde{\bf Q}_G\left( v-1, t \right) \; \widetilde{\bf Q}_G\left( v, t \right) \right] &\in\;& \mathbb{R}_{+}^{6v} \label{eq:gatex}
\end{flalign}

Note that $\mathbbm{I} \left[G^t\left(F\right) = j \right]$ is a binary function that returns 1 if $F$ is in the $j$th cluster and 0 otherwise, as determined by $G^t$. Now, we compute $X\left(t\right)$, a matrix representing $\mathbf{x}$ for current and historical partitions. Each row in $X\left(t\right)$ corresponds to a partition, and columns represent concatenated, normalized statistics vectors (across features and time).

\begin{flalign}
X\left(t\right) &= \left[\mathbf{x}\left(1\right), \mathbf{x}\left(2\right), \hdots, \mathbf{x}\left(t-1\right), \mathbf{x}\left(t\right) \right] &\in\;& \mathbb{R}_{+}^{t \times 6v} \label{eq:gatebigx}
\end{flalign}

Note that the same clustering, i.e., $G^t$, must be applied to each partition, or it won't make sense to compare rows in $X\left(t\right)$.





\subsubsection{Alert Generation}
\label{sec:solution-alertgen}

Given $X\left(t\right) \in \mathbb{R}^{t \times 6v}$ from the previous step, compare rows to determine whether the data at time $t$ should trigger an alert. Intuitively, each row, i.e., $X_i\left(t\right)$, represents the quality of $D^{i}$. We fix some neighbor fraction $f$, and for each row $i \in t$, we compute the average distance from $X_t\left(t\right)$ to the closest $\lfloor f \times t \rfloor$ non-anomalous neighbors, or preceding rows, according to \Cref{eq:nn}:

\begin{flalign}
\mathbbm{I}_y \left(t\right) &= \begin{cases} 1 & \text{if } y^t = 1 \\ \infty & \text{otherwise}\end{cases}   &\nonumber \\
\mathbf{d}\left(t\right) &= \left[ \mathbbm{I}_y \left(i\right)  \cdot \left\lVert X_t\left(t\right) - X_i\left(t\right) \right\rVert_p \text{ for } i \in \left[1, \hdots, t-1 \right] \right] &\in\;& \mathbb{R}_{+}^{t-1}  &\nonumber\\
f_{\gate} \left(t\right) &= \frac{1}{\lfloor f \times t \rfloor} \sum \min_{\lfloor f \times t \rfloor} \mathbf{d}\left(t\right) &\in\;& \mathbb{R}_{+} \label{eq:nn}
\end{flalign}

We experimented with setting $p = 1$ (Manhattan distance) and $p = 2$ (Euclidean distance) for $p$ in \Cref{eq:nn}. Similar to the adapted methods described in \Cref{sec:solution-adapting}, we trigger an alert if $f_{\gate} \left(t\right)$ exceeds a threshold, which can be determined by keeping track of $f_{\gate}$ for a few timestamps and determining if the current value of $f_{\gate}$ is an outlier. Our alert threshold is interpretable: one can multiply a ``maximum allowed'' $z$-score by the number of clusters (i.e., $v$) to get an estimate of how close two rows in $X^{t}$ can be to each other. 


\subsubsection{Debugging: Feature Drill-Down}
\label{sec:solution-drilldown}

Although the drill-down component is not a step in our automated data validation approach, \gate \rev{is designed to} help practitioners debug \rev{alerts by drilling into relevant clusters' features}. First, we rank the clusters by their corresponding $\mathbf{\widetilde{Q}}_G$ in $\mathbf{x}$ (\Cref{eq:gatex}). Then, for top clusters, we identify the features in each cluster using $G^t$ and similarly rank features by the magnitude of their $\mathbf{\widetilde{Q}}_z$. The top-ranked features are the most anomalously behaving features, by definition of $z$-score (i.e., larger absolute value $z$-scores are more standard deviations away from the mean), and can be presented to ML engineers. Another useful debugging strategy is to compare $X_t \left(t\right)$ to anomalous partitions, or $X_i\left(t\right)$ for $i < t$ where $y^{i} = 1$. If  $X_t\left(t\right)$ is a seasonal anomaly, such as Thanksgiving and Christmas, the neighboring anomalous $X_i\left(t\right)$ may also correspond to a holiday, since each partition is independently summarized. Additional information can be added to alerts described in~\Cref{sec:solution-alertgen} to give context on holidays and other practical patterns (e.g., weekday vs weekend patterns). \rev{We give an anecdote of the drill-down component in \Cref{sec:expts-discussion-oncall}.}

\section{Case Study}
\label{sec:expts}
\begin{figure*}
     \begin{subfigure}[b]{0.48\linewidth}
         \centering
         \resizebox{\linewidth}{!}{\input{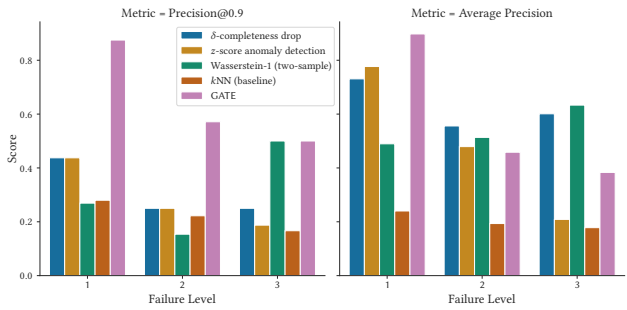}}
         \caption{Results for $\mathcal{D}_1$, where there are only a few failures. Most ML pipelines we've observed have similarly low failure counts.}
          \label{fig:d1allresultsgraph}
     \end{subfigure}
     \hfill
     \begin{subfigure}[b]{0.48\linewidth}
         \centering
         \resizebox{\linewidth}{!}{\input{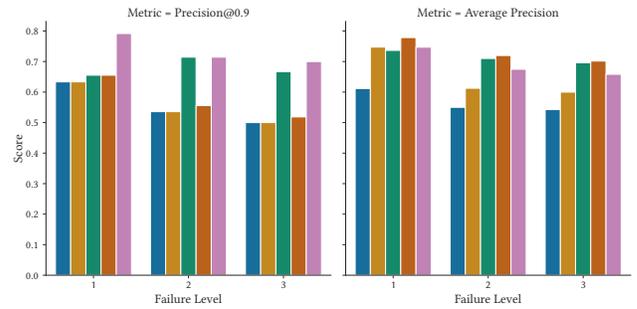}}
         \caption{Results for $\mathcal{D}_2$, where there are many failures (> 50\%). Our methods achieve higher precision@0.9 and comparable AP to the baseline \cite{Redyuk2021AutomatingDQ}.}
        \label{fig:d2allresultsgraph}
     \end{subfigure}
        \caption{\rev{Instagram case study} results across datasets and failure levels. \rev{The legend is shown in \Cref{fig:d1allresultsgraph}.}}
        \label{fig:allresultsgraph}
\end{figure*}

\rev{In this section, we report results from a case study on two of Instagram's ML pipelines. First, we present precision@0.9 and AP numbers for PS methods described in \Cref{sec:solution,sec:solution-gate}, as well as their runtimes. We choose parameter values based on business considerations (e.g., computing statistics with respect to rolling 7-day aggregations of data). Then, we discuss the usability of our techniques---characteristics of ML pipelines they are well-suited for and alert fatigue during on-call rotations.}

\subsection{\rev{Setup}}
\label{sec:expts-setup}


\subsubsection{Dataset Descriptions} Each of the two datasets in our case study included one month of partitions (i.e., $D^{1}, D^{2}, \hdots, D^{t}$ where $t \approx 30$) and tens of thousands of features (i.e., $F_1, F_2, \hdots, F_n$ where $n \geq 20,000$). We denote the specific datasets as $\mathcal{D}_1$ and $\mathcal{D}_2$ with feature sets $\mathcal{F}_1$ and $\mathcal{F}_2$ respectively. \rev{Since this case study was performed during the first author's internship, the datasets are drawn from the summertime.} $\mathcal{D}_1$'s minimum date partition is June 5, 2022 and maximum date partition is July 5, 2022. $\mathcal{D}_2$'s minimum date partition is July 15, 2022 and maximum date partition is August 15, 2022. Both datasets $\mathcal{D}_1$ and $\mathcal{D}_2$ share some (but not all) features---concretely, $\mathcal{F}_1 \ne \mathcal{F}_2$ and $\mathcal{F}_1 \cap \mathcal{F}_2 \ne \emptyset$. For privacy and legal reasons, we cannot disclose details of the features, but the datasets include a mix of integer, float, and categorical features.

As discussed in \Cref{sec:background-prodml}, models in the ML pipelines are frequently and automatically trained on fresh views of $\mathcal{D}$ and chained together to make final float-valued predictions, which represent the probability of a user clicking on a media recommendation (binary classification). \rev{We collaborated directly with on-call Instagram engineers to identify the types of ML performance drops they want the data validation system to flag. We found three categories:}

\begin{enumerate}
    \item \textbf{Increased model loss}: Whether the (normalized) cross-entropy loss on live predictions for an ML model (trained on $\mathcal{D}$) increased by some predefined threshold from its rolling 7-day average cross-entropy loss
    \item \textbf{Uncalibrated predictions:} Whether the error for a calibration function (fit on ML predictions to align them with true events) is larger than a predefined threshold
    \item \textbf{Label shift:} Whether the number of clicks on media in users' feeds decreased by some predefined threshold from its rolling 7-day average click rate
\end{enumerate}

For each $\mathcal{D}$, we unioned all occurrences of the three failure events to get ground truth labels $y^1, \hdots, y^t$ for the data validation methods. $\mathcal{D}_1$ and $\mathcal{D}_2$ have significantly different failure rates (i.e., fraction of positives), as shown in \Cref{tab:failurelevels}. $\mathcal{D}_2$ has an unusually high failure rate, so it will be easier for a method to achieve good precision and recall.

Finally, as mentioned in \Cref{sec:background}, ML teams at our organization care about the tunability of data validation methods, or the ability to adapt the method to different magnitudes of failures (i.e., model performance drops for the ML pipeline). For example, failures during days of major international news announcements may be more impactful than failures on a ``normal'' day in the year. To incorporate tunability into our \rev{case study}, we came up with different \emph{failure levels} for each dataset or ML pipeline. To derive a single failure level (FLs), we set thresholds for each failure definition, as shown in \Cref{tab:failurelevels}. We came up with three failure levels ($FL_1, FL_2, FL_3$) of increasing severity. Thresholds for each of the definitions increase such that the number of failures in $FL_i$ is $\geq$ the number of failures in $FL_{i+1}$. These FLs were hand-validated by an ML engineer to make sure positive ground-truth labels corresponded to actual model performance issues that the engineers needed to fix.


\begin{table}[h]
    \centering
    \footnotesize
    {\begin{tabular}{r|p{0.6\linewidth}|c|c}
\toprule
\textbf{FL} & \textbf{Definition} & $\bf \mathcal{D}_1$ & $\bf \mathcal{D}_2$ \\
\midrule
1   &  $(\uparrow \text{cross-entropy} \geq 0.02) \cup (\text{calibration error} \geq 0.3) \cup  (\text{label shift} \geq 0.3)$  &  0.33 &  0.59 \\ 
2   & $(\uparrow \text{cross-entropy} \geq 0.04) \cup (\text{calibration error} \geq 0.4) \cup  (\text{label shift} \geq 0.4)$  &  0.23 & 0.47 \\
3  & $(\uparrow \text{cross-entropy} \geq 0.06) \cup (\text{calibration error} \geq 0.5) \cup  (\text{label shift} \geq 0.5)$   &  0.17 & 0.44 \\
\bottomrule
\end{tabular}}
    \caption{Fraction of positive labels (i.e., failures) for each failure level (FL) in $\mathcal{D}_1$ and $\mathcal{D}_2$. }
    \label{tab:failurelevels}
\end{table}

\subsubsection{\rev{Data Validation Methods}} \rev{We evaluated \citet{Redyuk2021AutomatingDQ} as a baseline against our adaptations described in \Cref{sec:solution} ($\delta$-completeness drop, $z$-score anomaly detection, Wasserstein-1 two-sample statistical tests) and \gate.} \rev{The method proposed in \citet{Redyuk2021AutomatingDQ} creates a vector of statistics for each time step and performs a $k$-nearest neighbor algorithm against historical vectors to label the current time step's vector as anomalous or acceptable. We use the 7 statistics from the paper: completeness, approximate count of distinctive values, ratio of the most frequent value, maximum, mean, minimum, and standard deviation. For each feature, we compute these statistics and concatenate the results to get the larger vector used in the $k$-nearest neighbors algorithm. We call this method $k$NN (baseline) in \Cref{fig:allresultsgraph}.}

\subsubsection{\rev{Architectural Setup}} Due to privacy and legal considerations, we discuss only high-level details of the architectural setup.

Raw data (i.e., $\mathcal{D}$) for each dataset is stored in a data warehouse system. Upon arrival of a new partition, a PS job is launched: in this job, both ML model performance (e.g., accuracy or loss) and summary statistics are computed for that partition and logged to a summary store. The job's queries are written in PrestoSQL~\cite{sethi2019presto}. The statistics computed in the job cover all statistics in the data valiation methods---e.g., mean, standard deviation, minimum, maximum, completeness, eCDFs, frequent values, and more for each feature. As such, each data validation method reads from the same summary store to compare summaries and trigger alerts. The data validation methods are implemented in Python, Pandas, and \texttt{scikit-learn}. These jobs run for every partition summary, and evaluation metrics AP and precision@0.9 are computed with \texttt{scikit-learn} functions.

\subsection{\rev{Performance Results}}
\label{sec:expts-performance}

\begin{figure}
    \centering
    \resizebox{\linewidth}{!}{\input{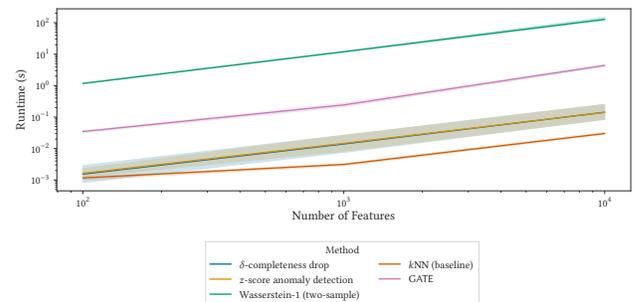}}
    \caption{Runtimes of computing a single partition's score for varying numbers of features, using three trials. \rev{The $\delta$-completion drop and $z$-score anomaly detection methods have similar runtimes.}}
    \label{fig:runtime}
\end{figure}

\rev{For $\mathcal{D}_1$, our best method gave a $2.9\times$ average improvement in precision@0.9 and $2.3\times$ average improvement in AP over the baseline. For $\mathcal{D}_2$, where there were more failures, our best method gave a $1.3\times$ improvement in precision@0.9 while maintaining $0.9\times$ the AP as the baseline.} We observed a reasonable runtime of approximately 5 seconds (50 seconds with the \rev{offline} decorrelation step described in \Cref{sec:solution-decorrelation}). In \Cref{fig:allresultsgraph}, we present the results for each method listed in \Cref{sec:expts-setup}. In \Cref{fig:runtime}, we show the runtimes to compute a single partition's score for each method, varying the number of features. \rev{We discuss how to select between methods for different ML pipeline settings in \Cref{sec:expts-discussion-bestmethods}.}

\subsubsection{Precision@0.9} In $\mathcal{D}_1$, \gate significantly outperformed other methods in precision@0.9. For FLs 1 and 2 in $\mathcal{D}_1$, \gate achieved approximately $2\times$ the precision@0.9 as other methods. For FL 3, \gate's precision@0.9 was matched by the Wasserstein-1 two-sample test. In $\mathcal{D}_2$, \gate achieved the highest precision@0.9 in all three FLs. The Wasserstein-1 method tied \gate's performance for FL 2 and came close for FL 3. \rev{In $\mathcal{D}_2$, where there were more failures, \gate's improvement in precision@0.9 was not as pronounced; nevertheless, it exhibited} the highest precision@0.9 across all failure levels. \rev{\gate has high precision due to its clustering step, which prevents correlated features from triggering many false positive alarms. The Wasserstein-1 two-sample test achieves a high precision when there are fewer failures (i.e., higher FLs) because in the case of a severe failure, a large fraction of records are corrupted, significantly perturbing entire distributions of features. Density measures (e.g., Wasserstein-1 distance) can capture this distributional shift more precisely than point statistics (e.g., mean), which aren't robust to outliers.}


\subsubsection{Average Precision (AP)} \rev{While AP is a less important metric than precision@0.9 for us, we found that our methods' APs were better than the baseline in $\mathcal{D}_1$ and nearly matched the baseline in $\mathcal{D}_2$.}  In $\mathcal{D}_2$, where there were significantly more failures, the baseline from \citet{Redyuk2021AutomatingDQ} achieved the highest AP for all FLs\rev{, but \gate achieved 90\% of that, and the Wasserstein-1 two-sample test achieved 97\% of that}. Still, we found our methods' APs acceptable in this setting, since most ML pipelines don't experience such a high failure rate (where it is easier to achieve good precision).

\subsubsection{Runtime} \label{sec:expts-overall-runtime} The baseline ($k$NN) is the fastest method because there is no time-based normalization to compute. The two-sample tests (highest AP) are the most computationally expensive because they require bootstrapping to yield $p$-values. They are not practical to execute for more than $O(10,000)$ features. \gate has the second-largest runtime, but its \rev{runtime is only a few seconds for $O(10,000)$ features.} Although \gate includes the Wasserstein-1 distance measure, it is faster than the two-sample tests because it does not bootstrap a $p$-value.


\subsection{\rev{Discussion}}
\label{sec:expts-discussion}

\rev{Here, we discuss methods to choose for different ML pipeline settings, \gate's feature drill-down component, and limitations of the baseline from \citet{Redyuk2021AutomatingDQ}.}

\rev{\subsubsection{Best PS Methods for Different Settings}\label{sec:expts-discussion-bestmethods} In our case study with Instagram ML pipelines and engineers and experience prototyping our data validation system with other teams, we learned that, if a data validation system needs to have high precision@0.9, different settings have different choices of best PS methods:}

\topic{Low failure rates} \rev{When pipelines have frequently-corrupted data, the corruptions are, by definition, less anomalous---rendering anomaly detection techniques like $z$-score computation useless. The Wasserstein-1 test (i.e., capturing the distribution of a feature) alone is the best single metric to use, but it is computationally expensive as it requires bootstrapping to get a $p$-value. \gate has higher precision, includes the Wasserstein-1 distance, and is orders of magnitude less computationally expensive (\Cref{fig:runtime}). Most of our ML pipelines had relatively low failure rates, making \gate a good overall choice. However, in pipelines with high failure rates, especially if AP matters more than precision@0.9, \citet{Redyuk2021AutomatingDQ}'s method might be preferable.}

\topic{Many correlated features} \rev{\gate significantly outperforms other methods in terms of mitigating false positives when ML pipelines have many correlated features because of \gate's clustering component (\Cref{sec:solution-decorrelation}), which only triggers an alert when an entire group of correlated features is anomalous. In our collaborations with other teams, we observed that many ML pipelines had hundreds, if not thousands, of correlated features. As it is the case in many other organizations~\cite{opmlinterview}, data scientists frequently create correlated features based on session data---for example, there might be binary features for watching a video after 1 second, 2 seconds, 5 seconds, and more. However, in pipelines with a few features, a simple, fast method like $\delta$-completion drop might be preferable.}


\topic{First week of deployment} \rev{We noticed that in the beginning of deployment, nearest-neighbor methods generated alerts in a seemingly random way. This was because there wasn't enough data to compare. Since the failure rate for most pipelines is typically less than or equal to 1 in 7, it took at least a week for \gate to start producing meaningful alerts. In the first week, the other methods (e.g., $\delta$-completeness drop) might be more useful to trigger alerts, since the alert criteria is typically more meaningful (e.g., drop of 30\%, $z$-score $\geq 3$, $p$-value $< 0.05$).} 

\topic{High-cardinality features} \rev{\gate worked well when there were features that could span a large distribution of values. In settings with only binary or low-cardinality (i.e., spanning only a few distinct values) features, data quality metrics like top-k values or number of unique values might, on their own, precisely flag anomalies. However, with real-valued or other high-cardinality features, such metrics might fluctuate for a feature even if the distribution of its values remains similar. In practice, many ML pipelines have a nontrivial amount of high-cardinality features: for example, user and content embeddings, which are float-valued, are often used in ML pipelines~\cite{pan2019social, cheng2016wide}.}

\subsubsection{On-Call Experiences}
\label{sec:expts-discussion-oncall}

\rev{In this case study, \gate reduced alert fatigue during on-call rotations by having fewer false positive alerts while maintaining acceptable recall requirements. Although \gate was primarily used to anticipate ML performance drops before they showed up in performance dashboards, the feature drill-down component (\Cref{sec:solution-drilldown}) could also be used during on-call rotations to diagnose why there was a performance drop. In one scenario, an ML pipeline's performance had significantly dropped, and several days had passed without identifying the root cause. Since there were more than 10,000 features in this model, most of the features had at least one anomalous data quality statistic, making it impossible to figure out the group of broken features. Upon looking at the feature names in the most anomalous cluster identified by \gate, it was clear that there was a sound-related bug, since many of the features were related to audio and had anomalous Wasserstein-1 values.}

\rev{\subsubsection{Limitations of the Baseline}\label{sec:expts-discussion-baselinedrawback} While the general-purpose data quality method proposed in \citet{Redyuk2021AutomatingDQ} achieved good AP in ML pipeline settings with high failure rates, we found that its precision@0.9 was too low for us to operationalize.} {First, it's not appropriate for time series settings because its normalization step (i.e., computing $\widetilde Q$) doesn't normalize partition summaries with respect to time.} To see how this could be bad, consider two partitions' summaries ${\bf h}{\left(1\right)}$ and ${\bf h}{\left(2\right)}$, containing two statistics (completeness and mean) and three features:

\begin{align*}
{\bf h}{\left(1\right)} &= \left[0.9, 13, 0.4, 1993, 0.8, 284448 \right] \\
{\bf h}{\left(2\right)} &= \left[0.9, 13, 0.4, 1993, 0.2, 300 \right]
\end{align*}

Each ${\bf h}$ is the concatenation of (fraction of null values, mean) tuples for each feature, and the only difference between ${\bf h}{\left(1\right)}$ and ${\bf h}{\left(2\right)}$is the null fraction and mean for the third feature. If we normalize according to~\citet{Redyuk2021AutomatingDQ}, or rescale each vector to be between 0 and 1, we get:

\begin{align*}
{\bf x}{\left(1\right)} &= \left[1.8 \times 10^{-5}, 4.4\times 10^{-4}, 0., 7.0\times 10^{-2}, 1.4\times 10^{-5}, 1. \right] \\
{\bf x}{\left(2\right)} &= \left[3.5\times 10^{-4}, 6.4\times 10^{-4}, 1.0\times 10^{-4}, 1., 0., 1.5\times 10^{-1} \right] 
\end{align*}

Post-normalization, the statistics that remained unchanged from $t = 1$ to $t=2$ now have differences of orders of magnitude---even though only one feature changed. This difference would confuse an alert trigger mechanism or draw an engineer's attention to the wrong features (e.g., features that didn't actually change). While this subtle flaw in normalization might go unnoticed in datasets with few types of errors or few columns, in practice, datasets for ML pipelines may have thousands or more features (i.e., columns), and there's almost always a different feature with one or more outlier values in each partition. Consequently, the results in \Cref{fig:allresultsgraph} show poor precision for this method. \rev{\gate avoids this problem by normalizing statistical measures with respect to time by using $z$-scores. Another issue we observed was that the $k$NN in \citet{Redyuk2021AutomatingDQ} treated each feature equally, which is not useful in the ML setting, where pipelines with many correlated features and varying feature importances.}

\section{Related Work}
\label{sec:related}
\gate combines insights from anomaly detection, \rev{clustering}, data cleaning, and data validation.

\topic{Anomaly Detection \rev{and Clustering} for Large Datasets} \citet{anomalydetectionsurvey} define anomaly detection as finding patterns in data that do not conform to expected behavior and enumerate challenges of adapting the notion of an anomaly as the underlying distribution of data changes over time.
\rev{While estimation models (e.g., $z$-score, Median Absolute Deviation) based on a threshold are commonly used to estimate point outliers~\cite{blazquez2021review, mehrang2015outlier}, our problem is more similar to the challenge of finding a subsequence of outliers. We draw inspration from dissimilarity approaches (e.g., $k$-nearest neighbors), which are commonly used to estimate outlier subsequences~\cite{ramaswamy, Angiulli2002FastOD, blazquez2021review}, but the challenge in our setting is to normalize inputs appropriately and handle correlated features, as discussed in~\Cref{sec:solution-gate}.  Several papers discuss anomaly detection in the light of high-dimensional data~\cite{Kamalov_2020, rousseeuw2011robust, huang2006network}. Methods like principal component analysis (PCA) produce low-rank representations of the data, which yield alerts interpretable at the tuple level, not the feature or column level (a requirement given by our engineers, as discussed in~\Cref{sec:background-existingdv}). \citet{blazquez2021review} survey clustering techniques for subsequences of data, but such methods don't work at Meta's scale. \citet{birch} introduce clustering large datasets using partitions of data, which we apply to our data validation problem. We summarize partitions and cluster the summaries.}


\topic{Data Cleaning for ML} It's well-known that data cleaning is important when constructing training sets for ML models~\cite{krishnan2016activeclean, wu2020complaint, gudivada2017data, chu2016data}. Most tools require manual input, such as verifying tuples predicted to be outliers, limiting their scalability~\cite{gdr, krishnan2016activeclean, wu2020complaint, katara}. The scalability concern is echoed by ~\citet{Stonebraker2018DataIT}, particularly in enterprises. ~\citet{dde} discuss four categories of data errors to consider when adopting cleaning approaches for large-scale datasets: outliers, duplicates, rule violations (e.g., violating uniqueness constraints), and pattern violation (e.g., failing type checks). The statistical measures used in our algorithm (e.g., completeness, frequent value, mean, standard deviation) span these categories of data errors. Overall, data cleaning methods are not exactly applicable to our setting because we are not interested in cleaning all corrupted tuples (our pipelines already have high-performing ML models). Rather, we simply wish to block the retraining of models on corrupted data. Moreover, at our scale, the challenge is in identifying corruptions that actually cause model performance drops, not any corruption.


\topic{Data Validation for ML Pipelines} Different data errors cause different magnitudes of ML model performance drops, motivating data validation for ML pipelines to catch egregious errors~\cite{automatinglargescale, breck,datalifecyclechallenges}. Several research projects and open-source software libraries aim to validate data flowing in out of ML pipelines. ~\citet{Biessmann2021AutomatedDV} discuss four dimensions of data validation: correctness, consistency, completeness, and statistical properties, and many data validation methods monitor these statistics over time. Existing work---such as Deequ~\cite{deequ}, TFX~\cite{baylor2017tfx}, DataSentinel~\cite{datasentinel}, and DaQL~\cite{Ehrlinger2019ADT}---defines constraints and libraries to monitor pipeline inputs and outputs (e.g., features and predictions), however users must declaratively specify constraint values (e.g., completeness bounds), which is not practical in our setting of thousands of datasets and tens of thousands of features in each dataset. Data ``linting'' tools typically perform type checks, duplicate detection, and outlier detection based on a fixed number of standard deviations away from the mean but do not tie directly to downstream ML model performance or anticipate model performance drops~\cite{datalinter, Ehrlinger2019ADT}. In a broader survey of data quality monitoring tools,~\citet{ehrlinger} find that most tools require a ``gold standard'' of data---which often doesn't exist in most production settings, including our setting. Additionally, adapting these tools to a continual setting requires manual fine-tuning and supervision, which we cannot do at our scale~\cite{ehrlinger}.~\citet{datavalexp} describe a case study of integrating data validation tools into their ML pipelines and find that, in practice, the integration process is painful, and most engineering teams ignore data validation alerts in their workflows. \rev{For instance, two-sample statistical tests based on differences between distributions (e.g., Wasserstein) commonly trigger many false positive alerts~\cite{massey1951kolmogorov, mullermetrics, vallender1974calculation}.}~\citet{datavalexp} say that the alerts did not provide actionable feedback (i.e., did not show which tuples or columns were broken) and required too much manual maintenance (e.g., declaratively specifying and fine-tuning constraints).~\citet{Redyuk2021AutomatingDQ} address the PS setting and formalize the problem of automating data validation for ``dynamic data ingestion'' (i.e., ingesting and validating new partitions of data for different applications). They introduce an automated $k$-nearest neighbors approach to validate partitions of data. Our method, inspired by~\citet{Redyuk2021AutomatingDQ}, is specific to the ML pipeline setting, where we want to anticipate downstream ML performance drops.

\topic{Evaluating Data Cleaning and Validation Methods} Prior work
on data cleaning and validation employ a variety of evaluation metrics~\cite{Redyuk2021AutomatingDQ, automatinglargescale, crowdsourcing, dataprepduplicate, mler, castelijns2019pywash, correlatedfusion}. \citet{dataprepduplicate} and \citet{correlatedfusion} choose the area under the P-R curve (auPRC). We use AP instead of auPRC: the auPRC estimate typically estimates area with the trapezoidal rule, which can be overly optimistic~\cite{boyd2013area}. Others choose the area under the receiver operating characteristic curve (auROC) metric, or the relationship between true positive rate (TPR) and false positive rate (FPR). Although the auROC metric provides a better picture of the stability of a method, it doesn't appropriately capture the tradeoff between precision and recall~\cite{lobo2008auc, prroc}. \citet{prroc} discuss how ROC curves overestimate a method's performance when there is a skew in the class distribution, because a large change in the number of false positives can yield a small change in the ROC's FPR. Most ML pipelines have relatively few numbers of failures (i.e., performance drops) compared to non-failures (i.e., performance remains the same); thus, there is a class imbalance or skew in the distribution of failures. 


\section{Conclusion}
\label{sec:conclusion}

In this paper, we discussed automatically validating data before downstream ML model performance drops occur. We gave an overview of production ML pipelines and requirements for ML-specific data validation at an industrial scale. We described the Partition Summarization (PS) approach to data validation, where summaries of timestamped partitions are compared to determine anomalous partitions. We introduced a general adaptation for existing data validation methods to the PS setting and \gate, our method that produces high-precision alerts without manual tuning from engineers. Finally, we discussed our learnings from implementing automatic data validation in production ML pipelines at Meta.

\bibliographystyle{ACM-Reference-Format}
\bibliography{main}
\techreport{\appendix

\clearpage
\section{Algorithms}
\label{sec:app-algos}
\begin{algorithm}
\caption{Clusters features into discrete groups.}
\label{algo:decorrelation}
\begin{algorithmic}[1]
\Function{Cluster}{$\mathcal{F}$, statistics}\Comment{Cluster $n$ features using $\kappa$ statistics functions}
\State $\mathbf{X} \gets \emptyset$ \Comment{Initialize covariances of statistics}
\State $v \gets -\infty$ \Comment{Initialize number of clusters}
\Statex

\For{statistic \textbf{in} statistics}
\For{$i, j$ \textbf{in} $[t] \times [n]$} \Comment{Compute statistic for each time step for each feature}
\State $Q_{i, j} \gets \mathrm{statistic}\left(F_j^{i}\right)$
\EndFor \Comment{$Q \in \mathbb{R}^{t \times n}$}
\Statex


\State $\Sigma_{\mathrm{statistic}} \gets \frac{1}{t-1} \sum\limits_{\substack{i=1}}^t \left(Q_i - \bar{Q} \right) \left(Q_i - \bar{Q} \right)^\top$\Comment{Covariance matrix $\Sigma_{\mathrm{statistic}} \in \mathbb{R}^{n \times n}$}

\State $\mathbf{\Lambda} \gets Q^{-1}\Sigma_{\mathrm{statistic}}Q$  \Comment{Sort eigendecomposition in decreasing order of eigenvalue}

\State $\mathbf{\lambda} \gets \frac{\mathrm{diag} \left( \mathbf{\Lambda} \right)}{\sqrt{\sum_{\lambda} \mathrm{diag} \left( \mathbf{\Lambda} \right)}}$ \Comment{Normalize eigenvalues to get $\mathbf{\lambda} \in \mathbb{R}^n$}

\Statex
\State $v_{\mathrm{statistic}} \gets \argmin_k 
\left( \sum\limits_{\substack{j=1}}^k \mathbf{\lambda}_j \right) > 0.95$ \Comment{Num elements of $\lambda$ to explain 95\% variance}
\State $v \gets \max \left(v, v_{\mathrm{statistic}} \right)$
\State $\mathbf{X} \gets \mathbf{X} \cup \{|\Sigma_{\mathrm{statistic}}|\}$

\EndFor

\Statex
\State $\mathbf{\bar{X}} \gets \mu(\mathbf{X})$ \Comment{$\mathbf{\bar{X}} \in \mathbb{R}^{n \times n} =$ similarities between features}

\State $\mathbf{G} \gets \mathrm{spectraldecomposition}\left(\mathbf{\bar{X}}, v \right)$ \Comment{Group $\mathbf{\bar{X}}$ into $v$ clusters}

\State \textbf{return} $\mathbf{G}$

\EndFunction
\end{algorithmic}
\end{algorithm}

\clearpage
\section{Results}
\label{sec:app-results}

\Cref{tab:deltacovdrop,tab:rawzetascore,tab:twosamp,tab:knn,tab:ourmethod,tab:alllevels} describe results for each method discussed in~\Cref{sec:expts}. 

\begin{table}[bp!]
\centering
    \begin{subtable}[h]{0.48\linewidth}
    \resizebox{\linewidth}{!}{%
    \begin{tabular}{lYcc}
\toprule
{} &       \textbf{auc} &  \textbf{AP} &  \textbf{precision@0.9} \\
$\delta$ &           &                &                \\
\midrule
\textcolor{black}{0.1}                     &  \textcolor{black}{0.819549} &       \textcolor{black}{0.730815} &       \textcolor{black}{0.437500} \\
0.2                     &  0.714286 &       0.709970 &       0.304348 \\
0.3                     &  0.714286 &       0.712619 &       0.280000 \\
0.4                     &  0.751880 &       0.670912 &       0.318182 \\
0.5                     &  0.616541 &       0.485421 &       0.280000 \\
0.6                     &  0.451128 &       0.380144 &       0.269231 \\
\bottomrule
\end{tabular}}

    \caption{Metrics for $\mathcal{D}_1$}
    \label{tab:deltacovdrop1}
    \end{subtable} \hfill \begin{subtable}[h]{0.48\linewidth}
    \resizebox{\linewidth}{!}{%
    \begin{tabular}{lYcc}
\toprule
{} &       \textbf{auc} &  \textbf{AP} &  \textbf{precision@0.9} \\
$\delta$ &           &                &                \\
\midrule
0.1 &  0.285088 &       0.517256 &       0.612903 \\
0.2 &  0.399123 &       0.611547 &       0.612903 \\
0.3 &  0.421053 &       0.575434 &       0.612903 \\
0.4 &  0.381579 &       0.580262 &       0.612903 \\
0.5 &  0.456140 &       0.609675 &       0.612903 \\
\textcolor{black}{0.6} &  \textcolor{black}{0.342105} &       \textcolor{black}{0.528449} &       \textcolor{black}{0.633333} \\
\bottomrule
\end{tabular}}

    \caption{Metrics for $\mathcal{D}_2$}
    \label{tab:deltacovdrop2}
    \end{subtable}
    
    \caption{Average precision (AP) and precision@0.9 for various $\delta$-coverage drops.}
    \label{tab:deltacovdrop}
\end{table}

\begin{table}[bp!] 
    \centering
    \begin{subtable}[h]{0.48\linewidth}
    \resizebox{\linewidth}{!}{\begin{tabular}{lYcc}
\toprule
{} &       \textbf{auc} &  \textbf{AP} &  \textbf{precision@0.9} \\
$z$ &           &                &                \\
\midrule
\textcolor{black}{-9}     &  \textcolor{black}{0.849624} &       \textcolor{black}{0.758968} &       \textcolor{black}{0.411765} \\
-8     &  0.827068 &       0.725490 &       0.411765 \\
-7     &  0.819549 &       0.717179 &       0.437500 \\
-6     &  0.781955 &       0.695535 &       0.388889 \\
-5     &  0.751880 &       0.604195 &       0.388889 \\
-4     &  0.654135 &       0.500470 &       0.291667 \\
-3     &  0.729323 &       0.529920 &       0.350000 \\
-2     &  0.669173 &       0.614648 &       0.291667 \\
\bottomrule
\end{tabular}}
\caption{Metrics for $\mathcal{D}_1$}
    \label{tab:rawzetascore1}
    \end{subtable} \hfill \begin{subtable}[h]{0.48\linewidth}
    \resizebox{\linewidth}{!}{%
    \begin{tabular}{lYcc}
\toprule
{} &       \textbf{auc} &  \textbf{AP} &  \textbf{precision@0.9} \\
$\delta$ &           &                &                \\
\midrule
-9 &  0.458333 &       0.587665 &       0.612903 \\
\textcolor{black}{-8} &  \textcolor{black}{0.462719} &       \textcolor{black}{0.600762} &       \textcolor{black}{0.612903} \\
-7 &  0.451754 &       0.590456 &       0.612903 \\
-6 &  0.421053 &       0.617432 &       0.612903 \\
-5 &  0.368421 &       0.593311 &       0.633333 \\
-4 &  0.412281 &       0.642711 &       0.612903 \\
-3 &  0.460526 &       0.664558 &       0.612903 \\
-2 &  0.535088 &       0.697619 &       0.612903 \\
\bottomrule
\end{tabular}}

    \caption{Metrics for $\mathcal{D}_2$}
    \label{tab:rawzetascore2}
    \end{subtable}

    \caption{Average precision (AP) and precision@0.9 for various $z$-score cutoffs. Metrics stay the same beyond $z = -9$.}
    \label{tab:rawzetascore}
\end{table}

\begin{table}[bp!]
\centering
    \begin{subtable}[h]{\linewidth}
    \resizebox{\linewidth}{!}{\begin{tabular}{lYY|cc|cc}
\toprule
{} & \textbf{auc} & something & \multicolumn{2}{c|}{\textbf{AP}} & \multicolumn{2}{c}{\textbf{precision@0.9}} \\
{} &      $p \leq 0.01$ &      $p \leq 0.05$ &          $p \leq 0.01$ &      $p \leq 0.05$ &          $p \leq 0.01$ &      $p \leq 0.05$ \\
2-sample test             &           &           &               &     & &      \\
\midrule
DTS         &  0.624060 &  0.631579 &      0.471625 &  0.475196 &      0.269231 &  0.269231 \\
Kolmogorov-Smirnov          &  0.631579 &  0.646617 &      0.374830 &  0.405442 &      0.269231 &  0.269231\\
\textcolor{black}{Wasserstein-1} &  \textcolor{black}{0.631579} &  \textcolor{black}{0.654135} &     \textcolor{black}{0.479278} &  \textcolor{black}{0.489552} &      \textcolor{black}{0.269231} &  \textcolor{black}{0.269231} \\
\bottomrule
\end{tabular}}
\caption{Metrics for $\mathcal{D}_1$}
    \label{tab:2samp1}
    \end{subtable} \hfill \begin{subtable}[h]{\linewidth}
    \resizebox{\linewidth}{!}{
\begin{tabular}{lYY|cc|cc}
\toprule
{} & something & something & \multicolumn{2}{c|}{\textbf{AP}} & \multicolumn{2}{c}{\textbf{precision@0.9}} \\
{} &      $p \leq 0.01$ &      $p \leq 0.05$ &          $p \leq 0.01$ &      $p \leq 0.05$ &          $p \leq 0.01$ &      $p \leq 0.05$ \\
2-sample test             &           &           &               &     & &      \\
\midrule
DTS &  0.649123 &  0.653509 &      0.696956 &  0.712853 &      0.612903 &  0.633333 \\
Kolmogorov-Smirnov &  0.482456 &  0.456140 &      0.594646 &  0.574086 &      0.612903 &  0.612903 \\
\textcolor{black}{Wasserstein-1} &  \textcolor{black}{0.692982} &  \textcolor{black}{0.701754} &      \textcolor{black}{0.737679} &  \textcolor{black}{0.736441} &      \textcolor{black}{0.655172} &  \textcolor{black}{0.655172} \\
\bottomrule
\end{tabular}}
    \caption{Metrics for $\mathcal{D}_2$}
    \label{tab:2samp2}
\end{subtable}
    \caption{Average precision (AP) and precision@0.9 for different 2-sample test statistics and $p$-value cutoffs. The DTS test statistic is a variance-weighted Wasserstein-1. The Wasserstein-1 metric is also known as the Earth-Mover's Distance.}
    \label{tab:twosamp}
\end{table}

\begin{table}[bp!]
\centering
    \begin{subtable}[h]{\linewidth}
    \centering
    {\begin{tabular}{lYY|cc|cc}
\toprule
{} & \multicolumn{2}{Y|}{\textbf{auc}} & \multicolumn{2}{c|}{\textbf{AP}} & \multicolumn{2}{c}{\textbf{precision@0.9}} \\
{} & oracle & unsupervised & oracle & unsupervised &  oracle & unsupervised \\
$k$ &                     &              &                     & & &             \\
\midrule
2             &            0.511278 &     0.300752 &            0.273940 &     0.235647 &            0.269231 &     0.269231 \\
3             &            0.545113 &     0.293233 &            0.310989 &     0.220618 &            0.269231 &     0.269231 \\
\textcolor{black}{4}             &            \textcolor{black}{0.575188} &     \textcolor{black}{0.353383} &            \textcolor{black}{0.317259} &     \textcolor{black}{0.240082} &            \textcolor{black}{0.269231} &    \textcolor{black}{0.280000} \\
5             &            0.500000 &     0.323308 &            0.266466 &     0.226088 &            0.269231 &     0.280000 \\
6             &            0.496241 &     0.308271 &            0.280943 &     0.219459 &            0.269231 &     0.280000 \\
7             &            0.447368 &     0.285714 &            0.247336 &     0.217364 &            0.269231 &     0.280000 \\
8             &            0.387218 &     0.278195 &            0.231693 &     0.215664 &            0.269231 &     0.280000 \\
9             &            0.334586 &     0.270677 &            0.236110 &     0.219775 &            0.269231 &     0.280000 \\
\bottomrule
\end{tabular}}
\caption{Metrics for $\mathcal{D}_1$}
    \label{tab:knn1}
    \end{subtable} \hfill \begin{subtable}[h]{\linewidth}
    \centering
    {\begin{tabular}{lYY|cc|cc}
\toprule
{} & \multicolumn{2}{Y|}{\textbf{auc}} & \multicolumn{2}{c|}{\textbf{AP}} & \multicolumn{2}{c}{\textbf{precision@0.9}} \\
{} & oracle & unsupervised & oracle & unsupervised &  oracle & unsupervised \\
$k$ &                     &              &                     & & &             \\
\midrule
2 &            0.414474 &     0.500000 &            0.576486 &     0.716186 &            0.612903 &     0.655172 \\
3 &            \textcolor{black}{0.576754} &     0.504386 &            0.673769 &     0.719567 &            0.612903 &     0.655172 \\
4 &            0.344298 &     0.517544 &            0.548290 &     0.732408 &            0.612903 &     0.655172 \\
5 &            0.344298 &     0.530702 &            0.560975 &     0.739248 &            0.612903 &     0.655172 \\
6 &            0.394737 &     0.548246 &            0.564589 &     0.749676 &            0.612903 &     0.655172 \\
7 &            0.392544 &     0.552632 &            0.560689 &     0.753588 &            0.612903 &     0.655172 \\
8 &            0.405702 &     0.570175 &            0.575454 &     0.766642 &            0.612903 &     0.655172 \\
9 &            0.513158 &     \textcolor{black}{0.600877} &            \textcolor{black}{0.629042} &     \textcolor{black}{0.778490} &            0.612903 &     0.655172 \\
\bottomrule
\end{tabular}}
    \caption{Metrics for $\mathcal{D}_2$}
    \label{tab:knn2}
\end{subtable}
    \caption{Average precision (AP) and precision@0.9 for various values of $k$ neighbors. In the unsupervised method (taken from the paper), for each datestamp, we average the distance to the $k$ closest previous neighbors as the outlier score. The highest scores indicate outliers, yielding a max-min objective. In the supervised method (an oracle we construct), we average the failure labels of the $k$ neighbors (across all time) to get the outlier score. Note that this method is an oracle because in practice, we won't have failure labels or data from the future.}
\label{tab:knn}
\end{table}

\begin{table}[bp!]
\centering
    \begin{subtable}[h]{\linewidth}
    \resizebox{\linewidth}{!}{\begin{tabular}{lYYY|ccc|ccc}
\toprule
 & \multicolumn{3}{Y|}{auc} & \multicolumn{3}{c|}{AP} & \multicolumn{3}{c}{precision@0.9} \\
 &         L1 &         L2 &         L3 &             L1 &         L2 &         L3 &             L1 &         L2 &         L3 \\
\midrule
Wasserstein-1 two-sample test       &  0.654135 &  0.681818 &  0.913043 &      0.489552 &  0.513462 &  0.633333 &      0.269231 &  0.153846 &  0.500000 \\
Novelty detection kNN  &  0.353383 &  0.488636 &  0.521739 &      0.240082 &  0.193423 &  0.178105 &      0.280000 &  0.222222 &  0.166667 \\
$\delta$-coverage drop &  0.819549 &  0.727273 &  0.739130 &      0.730815 &  0.555871 &  0.601010 &      0.437500 &  0.250000 &  0.250000 \\
$z$-score anomalies      &  0.864662 &  0.750000 &  0.652174 &      0.776681 &  0.479167 &  0.208824 &      0.437500 &  0.250000 &  0.187500 \\
\gate          &  0.962406 &  0.875000 &  0.869565 &      0.897619 &  0.458333 &  0.383333 &      0.875000 &  0.571429 &  0.500000 \\
\bottomrule
\end{tabular}}
\caption{Best metrics for $\mathcal{D}_1$}
    \label{tab:alllevels1}
    \end{subtable} \hfill 
    
\begin{subtable}[h]{\linewidth}
    \resizebox{\linewidth}{!}{\begin{tabular}{lYYY|ccc|ccc}
\toprule
 & \multicolumn{3}{Y|}{auc} & \multicolumn{3}{c|}{AP} & \multicolumn{3}{c}{precision@0.9} \\
 &         L1 &         L2 &         L3 &             L1 &         L2 &         L3 &             L1 &         L2 &         L3 \\
\midrule
Wasserstein-1 two-sample test       &  0.701754 &  0.812500 &  0.815126 &      0.736441 &  0.710054 &  0.696071 &      0.655172 &  0.714286 &  0.666667 \\
Novelty detection kNN  &  0.600877 &  0.637500 &  0.630252 &      0.778490 &  0.719735 &  0.701516 &      0.655172 &  0.555556 &  0.518519 \\
$\delta$-coverage drop &  0.456140 &  0.541667 &  0.567227 &      0.611547 &  0.550049 &  0.542434 &      0.633333 &  0.535714 &  0.500000 \\
$z$-score anomalies      &  0.618421 &  0.545833 &  0.554622 &      0.747156 &  0.612438 &  0.599779 &      0.633333 &  0.535714 &  0.500000 \\
\gate          &  0.754386 &  0.770833 &  0.781513 &      0.746975 &  0.674637 &  0.658178 &      0.791667 &  0.714286 &  0.700000 \\
\bottomrule
\end{tabular}}
    \caption{Best metrics for $\mathcal{D}_2$}
    \label{tab:alllevels2}
\end{subtable}
    \caption{Best results of each method for different failure levels.}
    \label{tab:alllevels}
\end{table}

\begin{table}[bp!]
\centering
    \begin{subtable}[h]{\linewidth}
    \centering
    {\begin{tabular}{lYY|cc|cc}
\toprule
{} & \multicolumn{2}{Y|}{\textbf{auc}} & \multicolumn{2}{c|}{\textbf{AP}} & \multicolumn{2}{c}{\textbf{precision@0.9}} \\
{} &         $p=1$ &         $p=2$ &             $p=1$ &         $p=2$ &             $p=1$ &         $p=2$ \\
\% history &           &           &               &           &               &           \\
\midrule
0.1 &  0.887218 &  0.932331 &      0.759066 &  0.840136 &      0.538462 &  0.700000 \\
0.2 &  0.924812 &  0.954887 &      0.807236 &  0.897619 &      0.636364 &  0.700000 \\
0.3 &  0.909774 &  0.954887 &      0.780612 &  0.897619 &      0.583333 &  0.700000 \\
0.4 &  0.932331 &  0.939850 &      0.816327 &  0.833333 &      0.700000 &  0.700000 \\
 0.5 &  0.924812 &  0.932331 &      0.816667 &  0.824242 &      0.583333 &  0.636364 \\
 0.6 &  0.947368 &  0.947368 &      0.844444 &  0.844444 &      0.777778 &  0.777778 \\
 \textcolor{black}{0.7} &  \textcolor{black}{0.962406} &   \textcolor{black}{0.962406} &       \textcolor{black}{0.873639} &   \textcolor{black}{0.873639} &       \textcolor{black}{0.875000} &   \textcolor{black}{0.875000} \\
0.8 &  0.954887 &  0.962406 &      0.859751 &  0.873639 &      0.777778 &  0.875000 \\
 0.9 &  0.962406 &  0.962406 &      0.873639 &  0.873639 &      0.875000 &  0.875000 \\
1.0 &  0.962406 &  0.962406 &      0.873639 &  0.873639 &      0.875000 &  0.875000 \\
\bottomrule
\end{tabular}}
\caption{Metrics for $\mathcal{D}_1$}
    \label{tab:ourmethod1}
    \end{subtable} \hfill \begin{subtable}[h]{\linewidth}
    \centering
    {\begin{tabular}{lYY|cc|cc}
\toprule
{} & \multicolumn{2}{Y|}{\textbf{auc}} & \multicolumn{2}{c|}{\textbf{AP}} & \multicolumn{2}{c}{\textbf{precision@0.9}} \\
{} &         $p=1$ &         $p=2$ &             $p=1$ &         $p=2$ &             $p=1$ &         $p=2$ \\
\% history &           &           &               &           &               &           \\
\midrule
0.1 &  0.622807 &  0.750000 &      0.673670 &  0.739080 &      0.678571 &  0.791667 \\
\textcolor{black}{0.2} &  \textcolor{black}{0.640351} &  \textcolor{black}{0.741228} &      \textcolor{black}{0.680040} &  \textcolor{black}{0.730462} &      \textcolor{black}{0.760000} &  \textcolor{black}{0.791667} \\
0.3 &  0.640351 &  0.723684 &      0.673461 &  0.734294 &      0.760000 &  0.791667 \\
0.4 &  0.561404 &  0.622807 &      0.610649 &  0.637687 &      0.730769 &  0.760000 \\
0.5 &  0.561404 &  0.583333 &      0.613855 &  0.620149 &      0.730769 &  0.760000 \\
0.6 &  0.548246 &  0.552632 &      0.627535 &  0.606947 &      0.678571 &  0.730769 \\
0.7 &  0.478070 &  0.504386 &      0.606334 &  0.581326 &      0.612903 &  0.703704 \\
0.8 &  0.385965 &  0.521930 &      0.572164 &  0.589734 &      0.612903 &  0.703704 \\
0.9 &  0.394737 &  0.289474 &      0.574556 &  0.496752 &      0.612903 &  0.612903 \\
1.0 &  0.394737 &  0.289474 &      0.574556 &  0.496752 &      0.612903 &  0.612903 \\
\bottomrule
\end{tabular}}
    \caption{Metrics for $\mathcal{D}_2$}
    \label{tab:ourmethod2}
\end{subtable}
    \caption{Average precision (AP) and precision@0.9 for \gate, based on the fraction of historical statistics vectors used in the nearest neighbors method. We compute distances between statistics vectors using the $\ell_p$ distance measure, for $p=1$ (Manhattan distance) and $p=2$ (Euclidean distance).}
    \label{tab:ourmethod}
\end{table}
}

\end{document}